\documentstyle[psfig]{mn2e}

\newcommand{\gsim}{\raisebox{-3.8pt}{$\;\stackrel{\textstyle >}{\sim}\;$}}
\newcommand{\lsim}{\raisebox{-3.8pt}{$\;\stackrel{\textstyle <}{\sim}\;$}}
\newcommand{\Msol}{$M_{\odot}$}

\newcommand{\etal}{\mbox{{\rm et~al.\ }}}
\newcommand{\ML}{\mbox{M$_*$/L}}

\newcommand{\MLi}{\mbox{M$_*$/L$_i$}}
\newcommand{\longsoft}{{\sf ls}}
\newcommand{\norsoft}{{\sf ns}}
\newcommand{\zeqzero}{$z=0$}

\title{The Tully--Fisher relation and its evolution with redshift 
in cosmological simulations of disc galaxy formation}

\author[Portinari \& Sommer--Larsen]{Laura Portinari$^1$ and 
                                     Jesper Sommer--Larsen$^2$ \\ 
       $^1$ Tuorla Observatory, University of Turku, V\"ais\"al\"antie 20,
            FIN-21500 Piikki\"o, Finland \\
       $^2$ Dark Cosmology Centre, Niels Bohr Institute, Juliane Maries Vej 
             30, DK-2100 Copenhagen, Denmark\\
E-mail: {\tt lporti@utu.fi, jslarsen@tac.dk}
}

\date{\tt Submitted: June 2006}

\pubyear{2006}
\begin{document}
\maketitle
\title{The TF relation and its redshift evolution in cosmological simulations}
%
%
\begin{abstract}
We present predictions on the evolution of the Tully--Fisher (TF) relation
with redshift, based on cosmological N-body/hydrodynamical simulations of 
disc galaxy formation and evolution. The simulations invoke
star formation and stellar feedback, chemical evolution with 
non-instantaneous recycling, metallicity dependent radiative cooling and
effects of a meta-galactic UV field, including simplified radiative transfer.
At $z=0$, the 
simulated and empirical TF relations are offset by about
0.4~magnitudes (1~$\sigma$) in the B and I bands. The origin
of these offsets is somewhat unclear, but it may not necessarily be a
problem of the simulations only.

As to evolution, we find a brightening
of the TF relation between $z=0$ and $z=1$ of about 0.85~mag in rest--frame 
B band, with a non-evolving slope. The brightening we predict is intermediate 
between the (still quite discrepant) observational estimates.

This evolution is primarily a luminosity effect, while the stellar mass
TF relation shows negligible evolution. The individual galaxies
do gain stellar mass between $z=1$ and $z=0$, by a 50--100\%; but they also 
correspondingly increase their characteristic circular speed. 
As a consequence, individually 
they mainly evolve {\it along} the stellar mass TF relation, while 
the relation as such does not show any significant evolution.
\end{abstract}


\section{Introduction}
\label{sect:introduction}
The Tully--Fisher (TF) relation is a most important distance indicator, but
also a key to understand the structure and evolution of disc galaxies.
Its slope is a standard test--bench for cosmologically motivated galaxy
formation models (Dalcanton, Spergel \& Summers 1997; Mo, Mao \& White 1998; 
Sommer--Larsen, G\"otz \& Portinari 2003, henceforth SGP03) and its absolute 
luminosity level 
is also a crucial constraint, suggesting low stellar mass--to--light ratios 
and  ``bottom--light'' Initial Mass Functions in disc galaxies 
(Portinari, Sommer--Larsen \& Tantalo 2004; henceforth PST04).

\begin{table*}
\label{tab:literature}
\caption{Overview of literature results on the evolution of the TF relation;
we indicate its brightening as $\Delta M_B = M_B (V_c,0)-M_B (V_c,z)$.
Then we report the redshift range and median redshift of the sample,
the number of galaxies and the reference paper.}
\begin{tabular}{c c c l p{4truecm}}
\hline
$\Delta M_B$ & $z$ range & $<z>$ & $N_{gal}$ & reference \\
\hline
\smallskip
1.5        & 0.15--0.35 & 0.25 & 24 & Rix \etal (1997) \\
\smallskip
1.5--2     & 0.25--0.45 & 0.35 & 12 & Simard \& Pritchet (1998) \\
\smallskip
0.4        &  0.1--1    & 0.5  & 16 & Vogt \etal (1996, 1997) \\

0.2        &  0.2--1    & 0.5  & $\sim$100 & Vogt (1999) \\
\smallskip
1.1        &  0.5--1.5  & 0.9  & 22 & Barden \etal (2003) \\
\smallskip
1.6 $z$    & 0.15--0.9  & 0.4  & 19 (field) & Milvang--Jensen \etal (2003) \\
$\sim 0.8^a$  & 0.05--1 & 0.45 & $\sim120$ & B\"ohm \etal (2004), \\
\smallskip
(1.22 $z$)$^a$ &         &      &           & B\"ohm \& Ziegler (2006)\\
\smallskip
0.3 & \multicolumn{3}{c}{revised analysis of B\"ohm \etal's data} & 
                                                  Kannappan \& Barton (2004) \\
\smallskip
1.0 $z$    &  0.1--1    & 0.33 & 89 & Bamford \etal (2006) \\
\smallskip
1.3 $z$    & 0.19--0.74 & 0.39 & 14 & Nakamura \etal (2006) \\
\smallskip
$\gsim 0$   &  0.4--0.75 & 0.53 & 11 (non-disturbed) & Flores \etal (2006) \\
\hline
\end{tabular}

{\footnotesize $^a$ For the sake homogeneous comparison to other data, 
we report in this table the B\"ohm \etal results for constant TF slope, \hfill
 
although the authors favour a slope evolution scenario to interpret their data 
(see text). \hfill}
\end{table*}

In the last decade, the increasing amount of observational data at intermediate
and high redshifts have allowed to study also the evolution of the TF relation,
as well as of other galaxy scaling relations. However, there is as of yet no 
convergence on the results (see the summary in Table~\ref{tab:literature}). 
Rix \etal (1997) and Simard \& Pritchet (1998)
argued for strong luminosity evolution of disc galaxies (1.5--2~mag in
rest--frame B band) already
at $z=0.3$. Conversely, Vogt \etal (1996, 1997) found at most moderate 
evolution ($\Delta M_B \lsim$0.4~mag) for a sample of about 16 spiral galaxies
at a median redshift of $z \sim 0.5$.
Later on, basing on a much larger sample of 100
galaxies between $0.2<z<1$, Vogt (1999) further reduced the estimated 
evolution to  $\Delta M_B \lsim$0.2~mag,
almost entirely driven by strong brightening of a few dwarf galaxies at 
$z>0.5$.
Most recently, from the analysis of the DEEP1 Groth Strip survey, 
Vogt (2006) reports no evolution at all (within $\pm$0.3~mag)
in the TF relation out to $z \sim$1, in the sense that all of the apparent 
evolution can be entirely accounted for by observational and selection effects.
Other groups instead find, out to $z \sim$1, a brightening up to 1~mag or more
(Rigopoulou \etal 2002; Barden \etal 2003; Milvang--Jensen \etal 2003; 
B\"ohm \etal 2004; Bamford, Arag\'on--Salamanca \& Milvang--Jensen 2006;
Nakamura \etal 2006). 
Significant brightening is also 
indicated by the line width -- luminosity relation 
at even higher redshifts: van Starkenburg \etal (2006) report a 2~mag 
offset for 3 galaxies at $z \sim$1.5. However, at
so early epochs, one major concern is how to interpret the 
kinematics of galaxies, which may not have settled onto a well developed
rotation curve comparable to those of local spirals.

Limiting our discussion to galaxies up to $z \sim$1, the discrepancy between
different results shows that these are plagued by observational 
difficulties and selection effects.
Smith \etal (2004) observed a disc galaxy at $z \sim$0.8 in integral field
spectroscopy, deducing for this one object a far more negligible brightening 
with respect to the local TF relation, than found by Barden \etal with 
long--slit spectroscopy;
this highlights the difficulties of measuring the kinematics 
of high redshift galaxies.
Kannappan \& Barton (2004) further suggest that
a significant contribution to the apparent evolution may be due to disturbed
kinematics biasing the measurements of rotation velocities in high
redshift samples, where kinematic anomalies are both more frequent (in the 
hierarchical clustering scenario) and more difficult to assess; 
kinematic disturbances might also contribute to the discrepancies between
results obtained from selecting large undisturbed discs (e.g.\ the Vogt paper 
series) and from other less selective samples.
Indeed recently, Flores \etal (2006) have used 3D integral field spectroscopy
to recover the velocity field of disc galaxies at $z=0.4-0.75$ and showed 
that the intermediate redshift TF relation of undisturbed rotating discs
shows much reduced scatter and evolution than a generic sample 
including perturbed or complex kinematics.

It seems therefore that the results on the TF evolution are still dominated
by observational difficulties, selection effects and systematics.
A recalibration of the zero--point of the local comparison TF relation 
by about 0.5~mag (Tully \& Pierce 2000 vs.\ Pierce \& Tully 1992; 
see Fig.~\ref{fig:TFz0}a) adds further confusion to the issue.

As to the slope of the TF relation, within the large scatter of the high 
redshift samples the slope is not very well constrained directly so the 
``minimal assumption'' is usually adopted of an evolution at constant slope, 
i.e.\ of a uniform luminosity offset. However, there are some 
indications that evolution might
be mass dependent, and that sub--$L_*$ objects drive the brightening
(Simard \& Prichet 1998; Vogt 1999); differential mass evolution 
might partly explain the discrepancies between the results from different
samples. In their recent VLT sample B\"ohm \etal (2004) and B\"ohm \& Ziegler
(2006) find a brightening of about 0.8~mag if a constant TF slope is imposed, 
but their data are better fit by a changing slope implying much stronger 
brightening for dwarf galaxies. Kannappan \& Barton (2004) suggest though 
that this differential behaviour might be
driven by kinematic anomalies, namely that the objects with stronger evolution
are likely kinematically disturbed rather than genuinely brighter;
comparing the B\"ohm \etal data to a low redshift sample with similar 
selection as the high redshift one, they revise the estimated TF evolution 
down to a modest $\Delta M_B$=0.3~mag, with no change in slope.
Bamford \etal (2006) further point out that a spurious change in slope 
may be due to intrinsic coupling between scatter in rotation speed, 
TF residuals and limited magnitude range of the sample; also Weiner \etal 
(2006) find no slope evolution in a revised analysis of the B\"ohm \etal
(2004) data. Finally, very recently Weiner \etal (2006), from a large sample
of TKRS/GOODS galaxies with $z$=0.4--1.2, argue for an evolution in TF slope
{\it in the opposite sense}, with brighter, massive galaxies fading more than
low mass galaxies. The issue of slope evolution remains quite controversial.

However quantitatively significant the luminosity evolution of the TF 
relation, there is some evidence that it must be largely a luminosity 
dimming effect of the aging stellar populations, while in terms of the 
stellar mass there is hardly any offset, at fixed $V_c$, between the TF 
relation at $z=0$ and at $z=1$ (Conselice \etal 2005; Flores \etal 2006).

In this paper we present our predictions on the evolution of the TF
relation with redshift, based on cosmological+hydrodynamical simulations 
of disc galaxy formation. 
In Section~\ref{sect:simulations} we describe our simulations and their 
analysis.
In Section~\ref{sect:TFz0} we discuss the theoretical TF relation at $z=0$
and the effects of numerical resolution and softening length on the results. 
In Section~\ref{sect:TFevolution} we illustrate the evolution 
of the TF relation as predicted by our simulations.
A special experiment with no infall of hot halo gas after $z=1$ is discussed 
in Section~\ref{sect:noinfall}.
Finally, Section~\ref{sect:conclusions} outlines our conclusions.

\section{The simulations}
\label{sect:simulations}
The code used for the simulations is a significantly improved version of
the TreeSPH code we used for our previous work on galaxy formation 
(SGP03). A similar version of the code has been used recently
to simulate clusters of galaxies, and a detailed description can be found 
in Romeo \etal (2006). Here we briefly mention its main features and the 
upgrades over the previous version of SGP03 --- see also Sommer-Larsen (2006).
\begin{enumerate}
\item
The basic equations are integrated by incorporating the ``conservative'' 
entropy equation solving scheme of Springel \& Hernquist (2002), which
improves the numerical accuracy in lower resolution regions.
\item
Cold high-density gas is turned into stars in a probabilistic way as
described in SGP03. In a star-formation event 
a SPH particle is converted fully into a star particle. Non-instantaneous 
recycling of gas and heavy elements is described through probabilistic 
``decay'' of star particles back to SPH particles as discussed by 
Lia \etal (2002a). In a decay event a star particle is converted fully 
into a SPH particle, so that the number of baryonic particles in the simulation
is conserved.
\item
Non-instantaneous chemical evolution tracing
10 elements (H, He, C, N, O, Mg, Si, S, Ca and Fe) has been incorporated
in the code following Lia \etal (2002a,b); the algorithm includes 
supernov\ae\ of type II and type Ia, and mass loss and chemical enrichment
from stars of all masses.
For the simulations presented in this paper, we adopt the Initial Mass 
Function (IMF) of Kroupa (1998), derived for field stars in the Solar 
Neighbouhood; this IMF well reproduces the chemo--photometric properties 
of the Milky Way 
(e.g.\ Boissier \& Prantzos 1999) and is ``bottom--light'', which aids in
reproducing the luminosity level of the TF relation (PST04). Two experiments
have been run with the Salpeter IMF for comparison.
\item 
Atomic radiative cooling is implemented, depending both on the 
metallicity of the gas (Sutherland \& Dopita 1993) and on the meta--galactic 
UV field, modelled after Haardt \& Madau (1996). Moreover, a simplified 
treatment of radiative transfer, switching off the
UV field where the gas becomes optically thick to Lyman limit photons on
scales of $\sim$ 1~kpc, is invoked.
 \item
Star-burst driven winds are incorporated in the simulations
at early epochs, as strong early feedback is crucial to largely overcome 
the angular momentum problem (SGP03). A burst of star 
formation is modelled in the same way as in SGP03: when a star particle 
is formed, further self-propagating star formation is triggered in the 
surroundings; the energy from the resulting, correlated
SNII explosions is released initially into the interstellar medium 
as thermal energy, and gas cooling is locally halted to reproduce the 
adiabatic super--shell expansion phase; a fraction of the supplied energy 
is subsequently converted (by the hydro code itself) into kinetic energy 
of the resulting expanding super--winds and/or shells.
The super--shell expansion also drives the dispersion of the metals produced
by type~II supernov\ae\ (while metals produced on longer timescales are 
restituted to the gaseous phase by the ``decay'' of the corresponding 
star particles, see point 2 above).

At later epochs, only a fraction (typically, 20\%) of the stars induce 
efficient feedback, and star formation is no longer self--propagating
so that no strong starbursts are triggered by correlated SN explosions.
This allows the smooth settling of the disc (see SGP03 for all details).
\end{enumerate}
The galaxies were drawn and re-simulated from a $10 h^{-1}$~Mpc 
box-length dark matter (DM)-only 
cosmological simulation, based on the ``standard'' flat $\Lambda$ 
Cold Dark Matter cosmological model ($h=0.65$, $\Omega_0=0.3$, 
$\sigma_8=1.0$); our choice of $h$ and $\sigma_8$ is slightly
different from presently more popular values (0.7 and 0.9 respectively),
but this has little impact on the Tully-Fisher locus (see 
\S~\ref{sect:diffUniv}).
When re-simulating with the hydro-code, baryonic 
particles were ``added" to the original DM ones, which were
split according to an adopted baryon fraction $f_b=0.15$. 
The gravity softening lengths were fixed in physical coordinates from $z$=6
to $z$=0 and in comoving coordinates at earlier times.

Our model galaxies are typically run with a resolution of 
$m_{SPH}=m_*=1.1 \times 10^6$~\Msol, $m_{DM}=6.5 \times 10^6$~\Msol\ and 
$\epsilon_{SPH}=0.6$~kpc (``normal resolution'' henceforth);
a few cases were run at ``lower resolution'', with 
$m_{SPH}=m_*=9 \times 10^6$~\Msol, $m_{DM}=5.2 \times 10^7$~\Msol\ and a
gravitational softening length $\epsilon_{SPH}=\epsilon_*=1.2$~kpc, and some 
more at ``high resolution'' with
$m_{SPH}=m_*=1.4 \times 10^5$~\Msol\ and $\epsilon_{SPH}=0.3$~kpc. These
are the characteristics of ``normal softening length'' (\norsoft) simulations.
For smaller galaxies ($V_c \lsim$150~km/sec), we run some normal and high
resolution simulations with ``long softening length'' (\longsoft), where 
$\epsilon_{SPH}=1.2$~kpc and $\epsilon_{SPH}=0.6$~kpc respectively.

We show below that the predicted locus and evolution 
of the TF relation is largely robust to changes in the numerical
resolution.
Images of the simulated galaxies are available at 
{\sf http://www.tac.dk/$\sim$jslarsen}.

\begin{figure*}
\leavevmode
\psfig{file=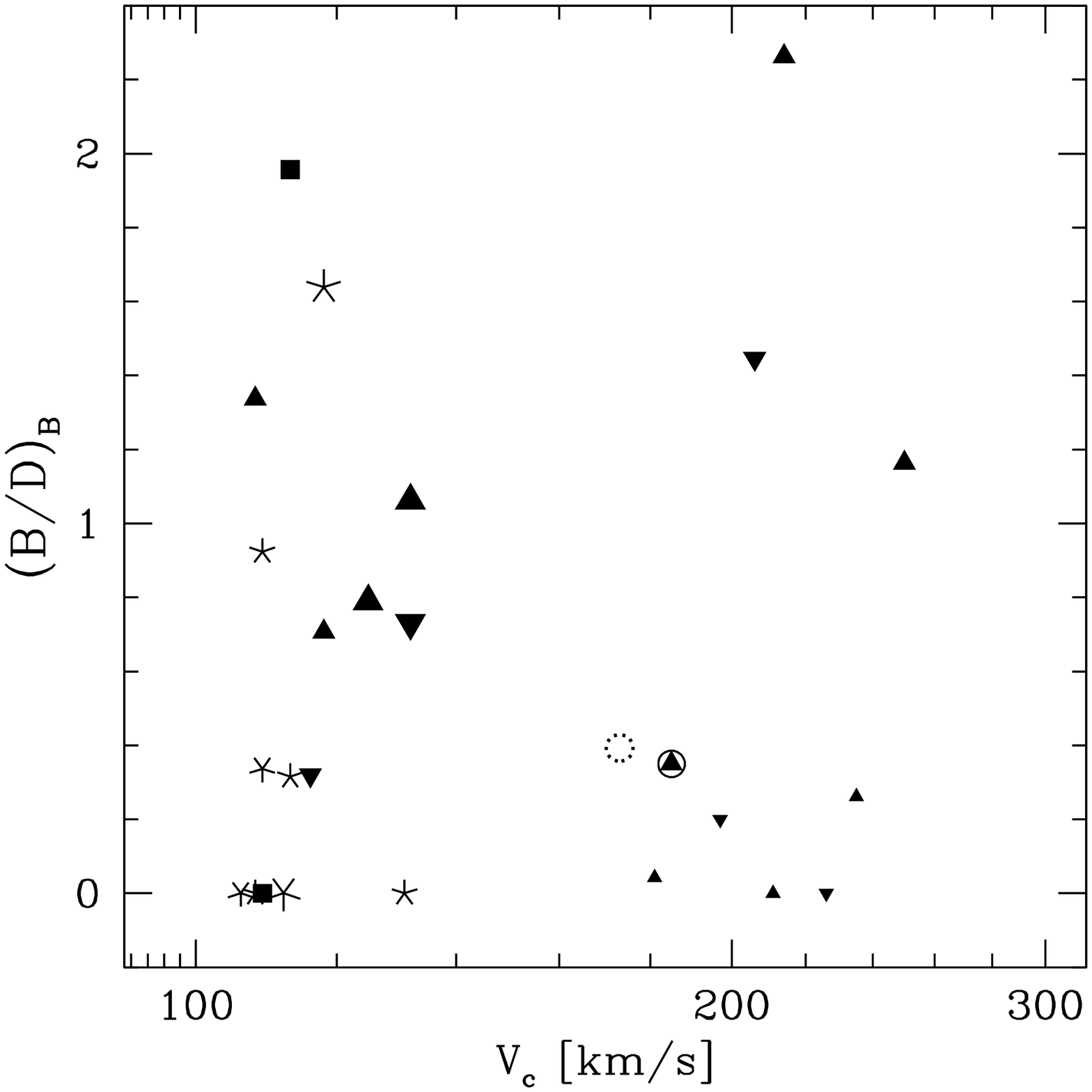,width=0.41\textwidth}
\psfig{file=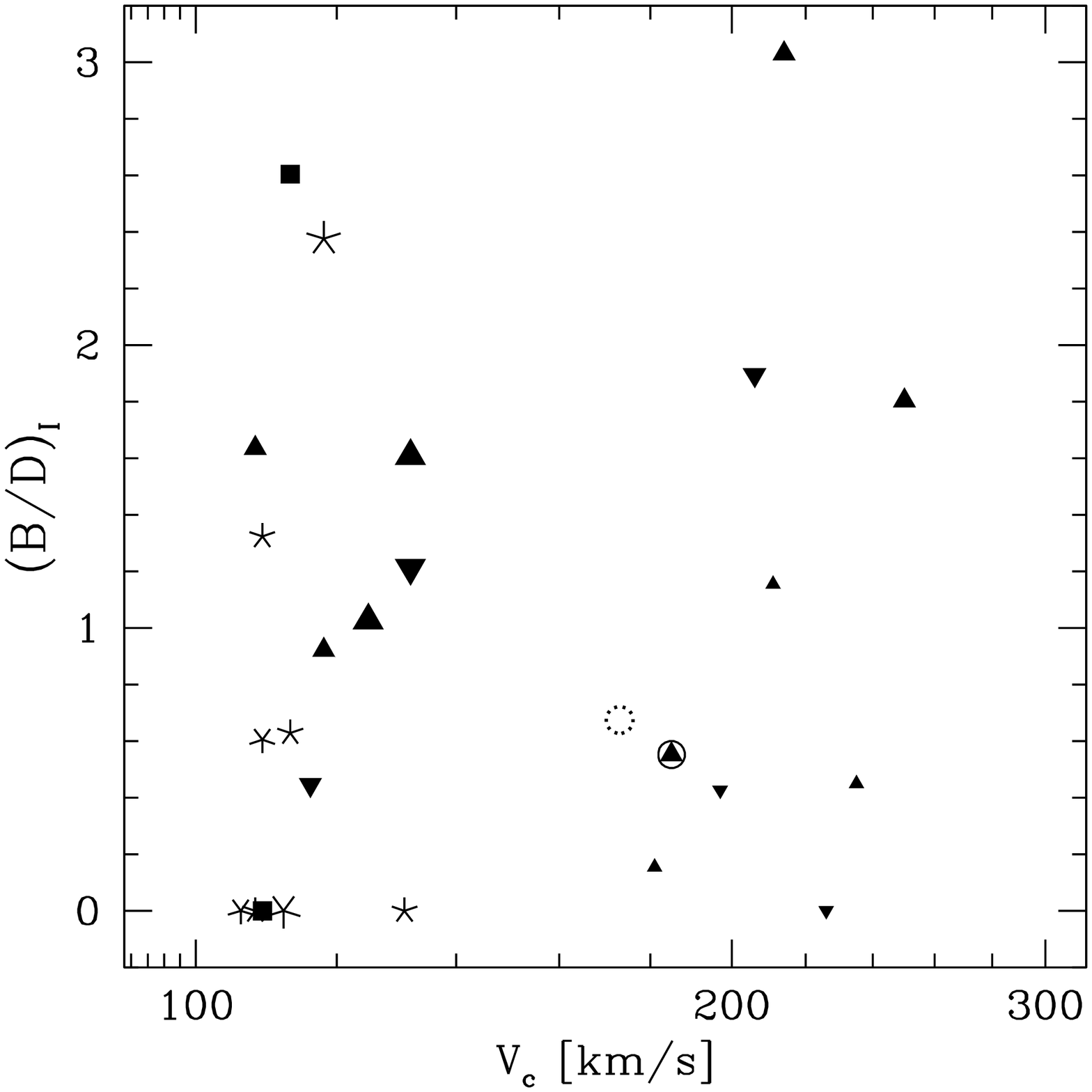,width=0.41\textwidth}

\leavevmode
\psfig{file=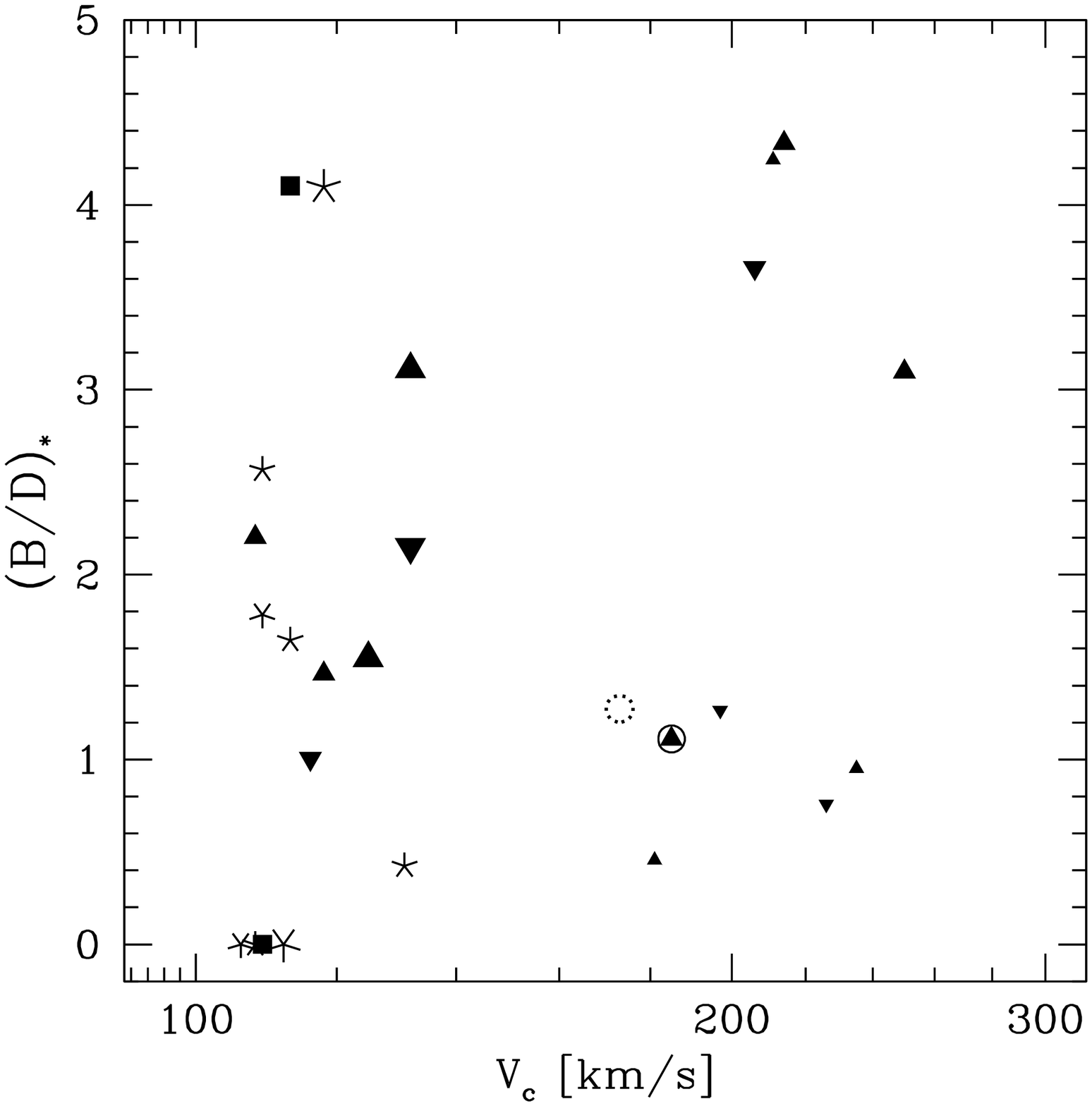,width=0.41\textwidth}
\psfig{file=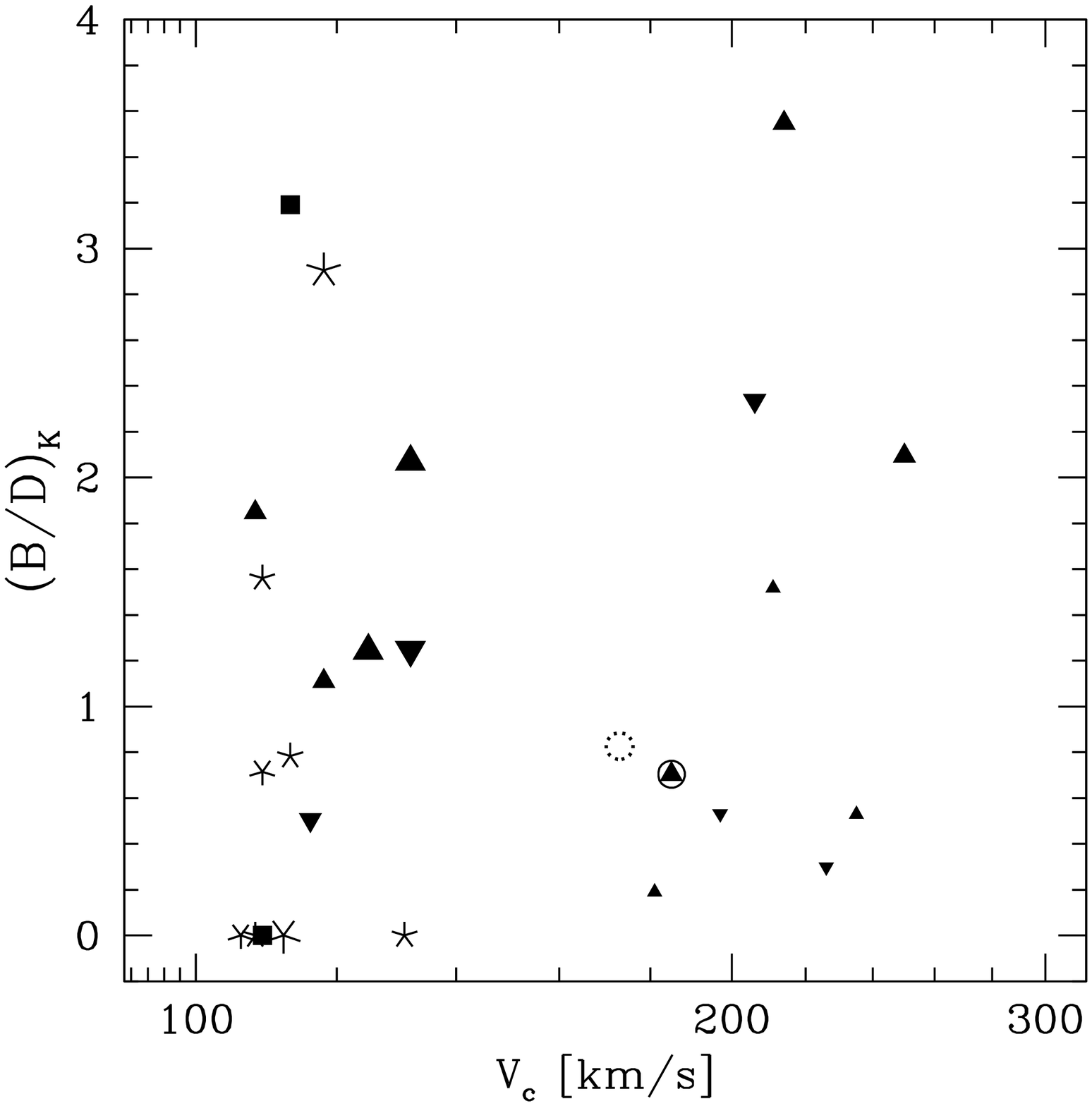,width=0.41\textwidth}
\caption{Bulge--to--disc ratio of simulated galaxies; clockwise from top left
panel: in B, I, K band and in the actual stellar mass profile. 
The size of the dots increases with the numerical resolution of the 
corresponding simulation; ``pairs'' of triangles with the same orientation, 
pointing upward or downward (and same colour, in the colour version 
of the figure) mark simulations of the same object (or cosmological halo) 
resimulated at two different resolution levels. Asterisks show simulations 
run in the ``long softening'' (\longsoft) mode. 
{\it Dotted circle}: the special ``no late infall'' experiment discussed 
in Section~\protect{\ref{sect:noinfall}}; the solid circle encircles 
the corresponding ``normal infall'' simulation.}
\label{fig:B/D}
\end{figure*}

\begin{figure*}
\leavevmode
\psfig{file=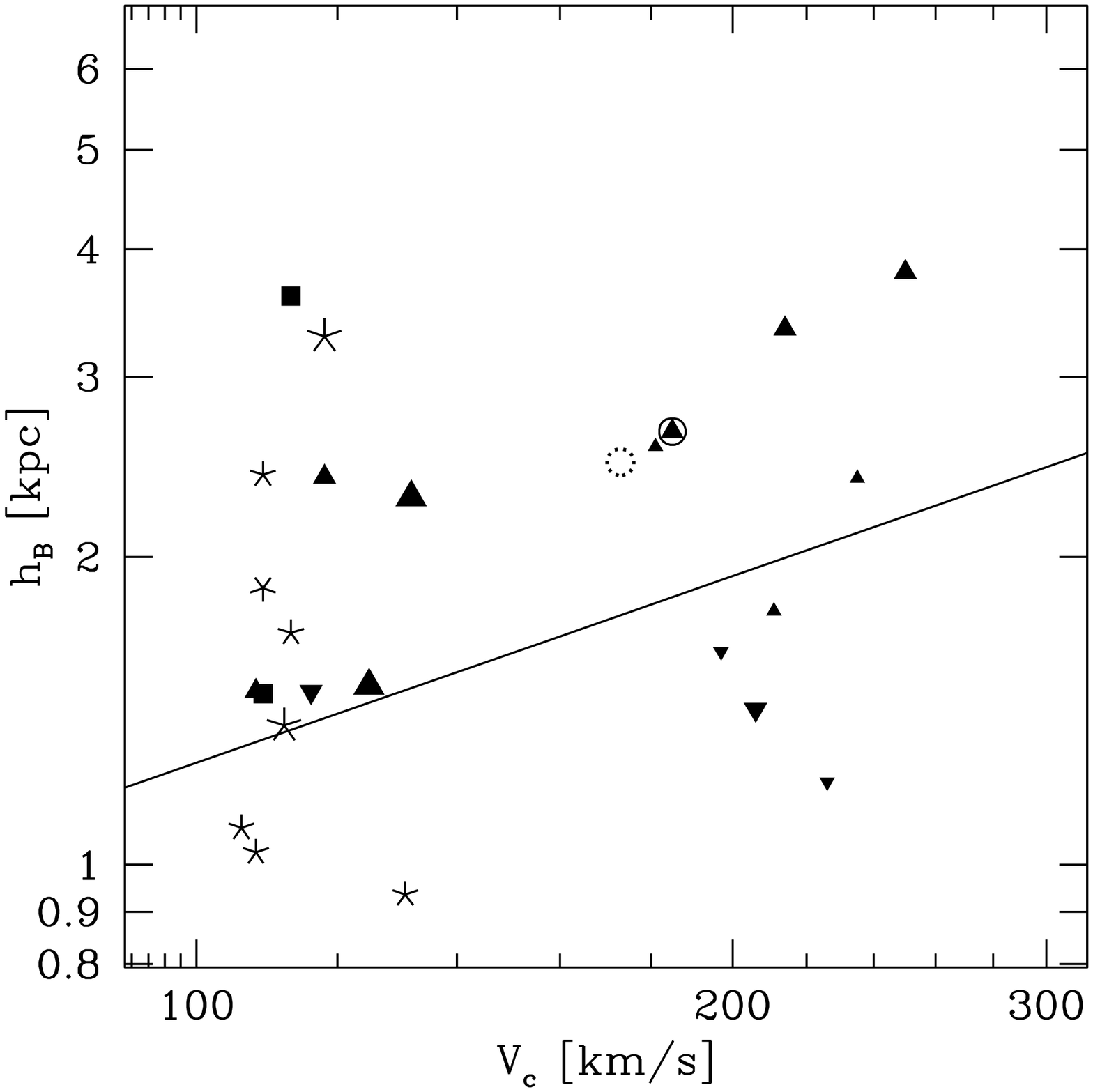,width=0.41\textwidth}
\psfig{file=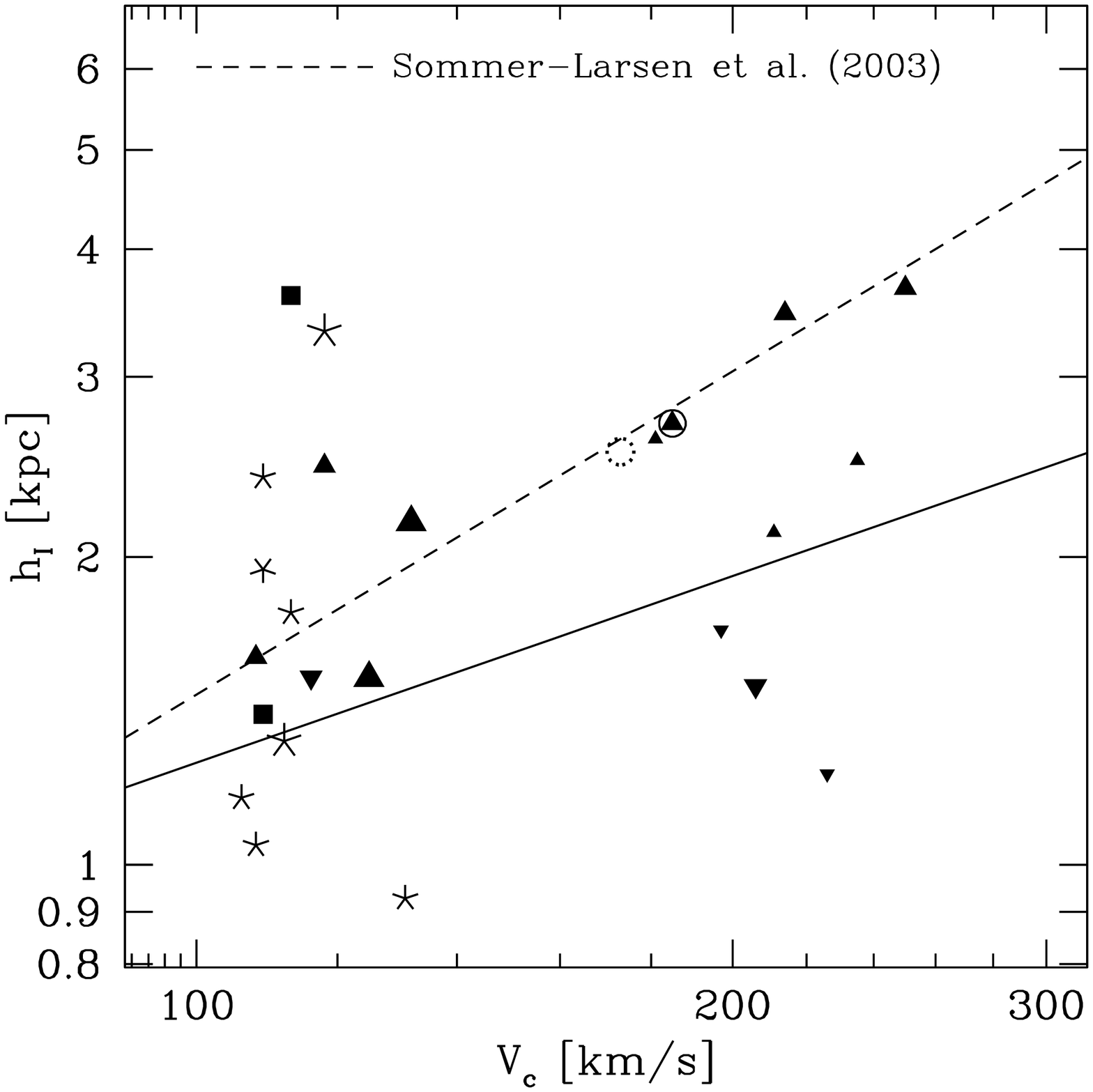,width=0.41\textwidth}

\leavevmode
\psfig{file=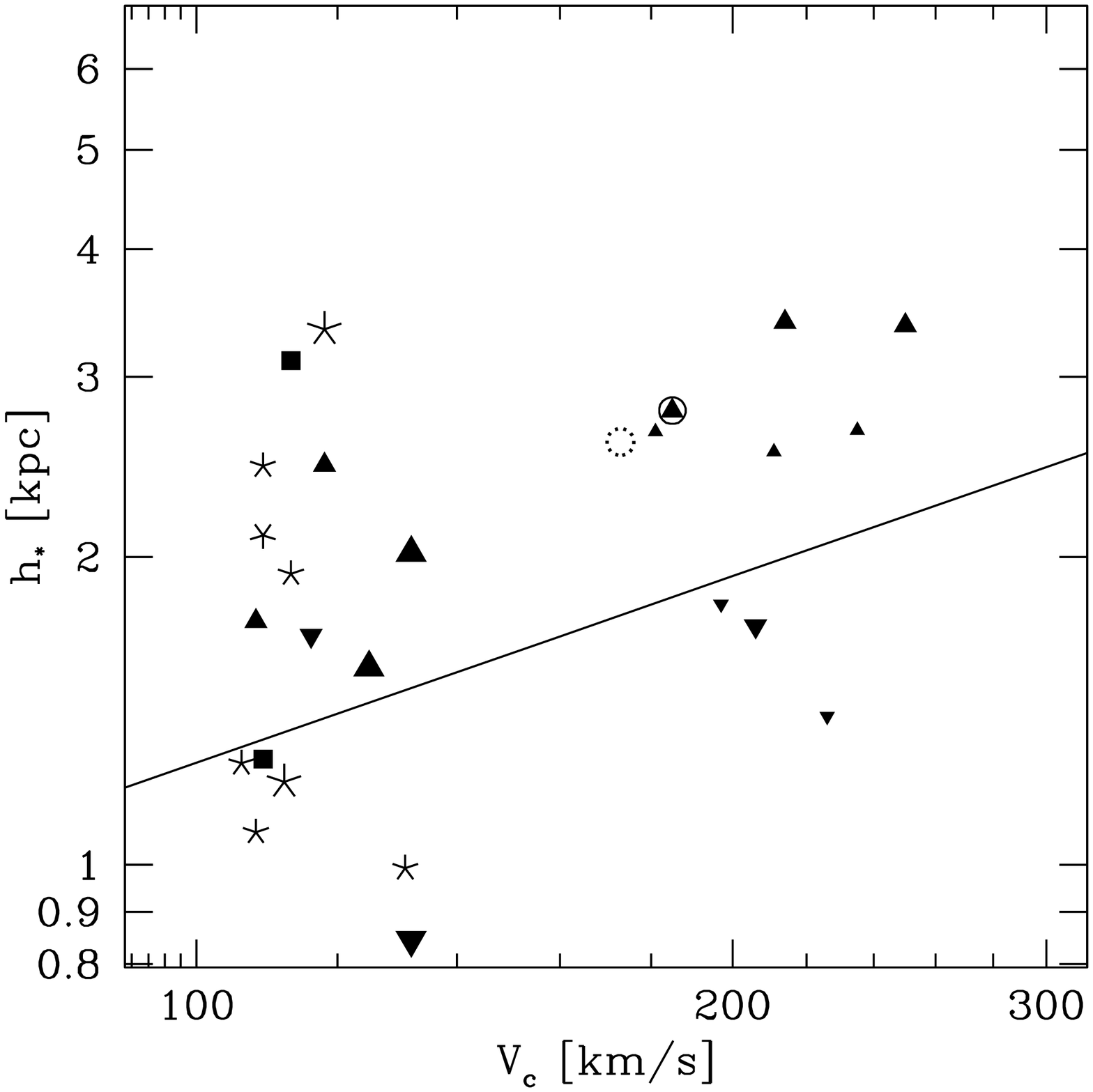,width=0.41\textwidth}
\psfig{file=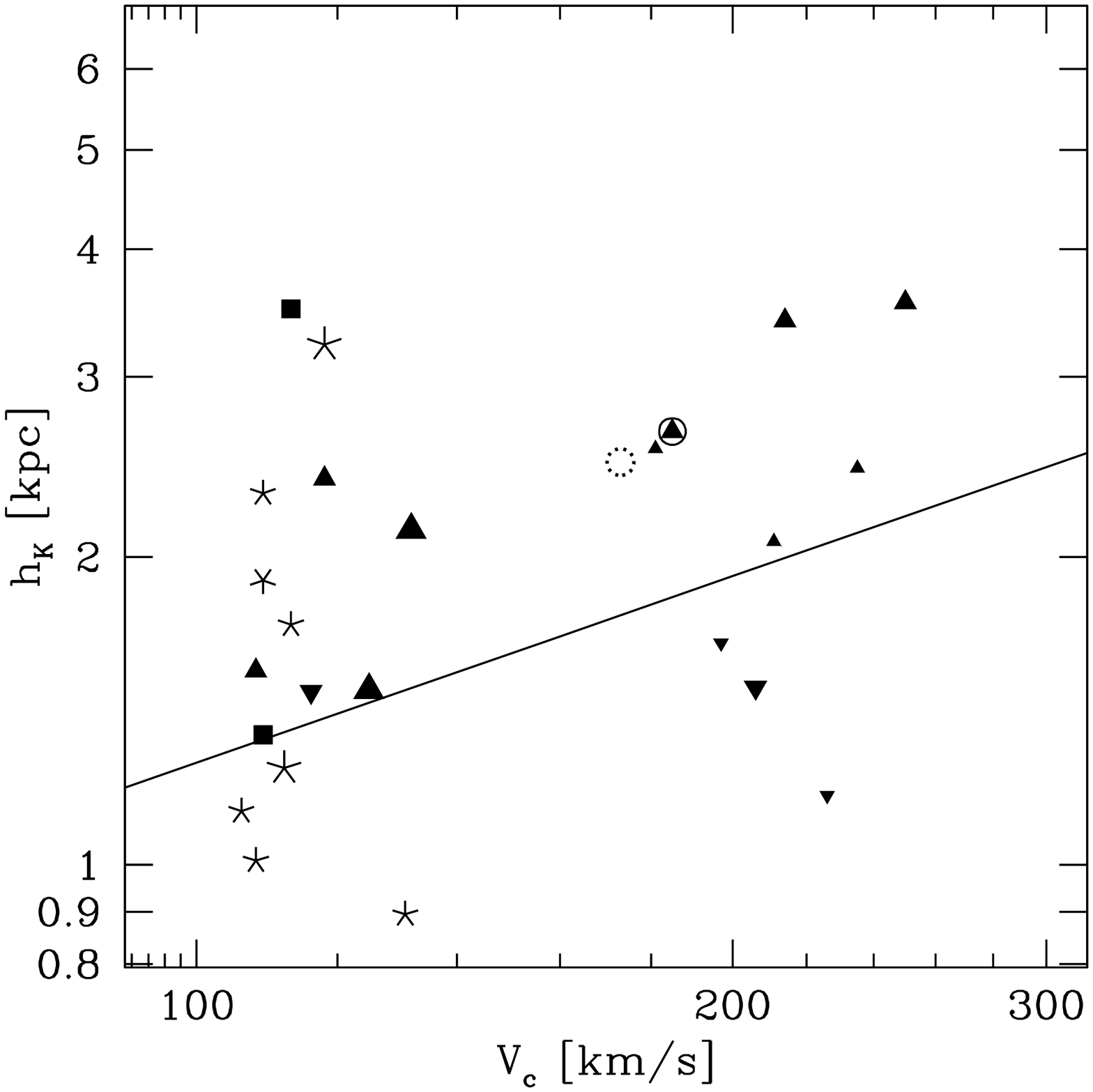,width=0.41\textwidth}
\caption{Disc scalelengths of simulated galaxies in B, I, K band and
in the actual stellar mass profile. Symbols as in Fig.~\protect{\ref{fig:B/D}}.
The {\it solid lines} show linear fits to the disc scalelengths, including 
only simulated galaxies with a realistic (B/D)$_I \leq 0.6$. 
In the I band panel (top right), the observational relation by
Sommer--Larsen (2003; see also Sommer--Larsen \& Dolgov 2001) is also shown
for comparison.}
\label{fig:scalelengths}
\end{figure*}

\subsection{The analysis}
\label{sect:analysis}
We determine the disc plane of a simulated galaxy from the direction of the
angular momentum vector of the star particles around the center of mass
of the system. 
Each star particle in the simulation represents a Single Stellar Population
(SSP) of given age and metallicity, whose luminosity in various bands is
straightforwardly computed from the SSPs of PST04 (see also Section~2.2 of
Romeo, Portinari \& Sommer--Larsen 2005). Hence we compute the surface 
brightness profile of each galaxy on the disc plane; a thickness of 3~kpc
is typically assumed for the disc stars, sometimes thinner or thicker
(between 2~kpc and 5~kpc) as judged from visual inspection of the simulation.
This limit ensures that 90--100\% of the stars are included out to the
very outskirts of the disc; in the inner disc regions, within 2 scalelengths
where most of the stellar mass resides, the scaleheights are significantly 
smaller and more realistic (0.5--1~kpc for Milky Way--sized simulated 
galaxies; this is the vertical resolution achieved with the softening lengths
typical of present--day fully cosmological simulations; cf.\ also Abadi
\etal 2003; Governato \etal 2006). Since the Tully-Fisher is a global relation,
the detailed vertical 
structure of our discs is not so crucial provided we make sure to include 
the bulk of the disc luminosity; indeed we verified that our results are 
robust with respect to the assumed thickness (e.g.,\ 3 vs.\ 2 kpc).
The large scaleheights do not have either a major impact on the effectiveness
of the feedback 
process, which in our simulations is strong and crucial only in the early, 
pre-galactic starbursts, while it is much milder for the main part of bulge 
and disc formation. The limited resolution on the vertical structure of discs 
is thus not a concern for our basic picture of disc galaxy formation.

The one--dimensional radial profiles of stellar surface mass density and 
UBVRIJHK surface brightness are each decomposed into an exponential disc and
a (de Vaucouleurs or exponential) bulge component by means of a
Leverberg--Marquardt fitting algorithm (Press \etal 1992). Any bar component 
is included in the bulge and not treated separately. The fit is
performed over the radial extent of the simulated discs as judged by visual
inspection; typical disc radii range from 8--10~kpc for small galaxies,
to 12--15~kpc (sometimes out to 20~kpc) for larger ones. The decomposition is
in any case quite robust to the assumed disc edge.
The decomposition with an exponential bulge profile provides better 
fits in general, so we will discuss this case
only in the following. Assuming a de Vaucouleur bulge obviously results in
larger bulge--to--disc (B/D) ratios but, although the central disc
brightness is correspondingly lower, neither the derived disc scalelengths 
nor our results on circular velocities and the
TF relations, discussed below, are much affected.

Fig.~\ref{fig:B/D} and Fig.~\ref{fig:scalelengths} show the B/D ratios
(with exponential bulge component) and the disc scalelengths respectively,
obtained for the stellar mass density profiles and the BIK surface 
brightness profiles of our galaxies. As expected, B/D ratios systematically
increase going from blue to red bands, and to the actual stellar mass profile.
Notice though that we also obtain a few bulgeless objects --- although 
some of the apparent ``pure discs'' in B band are no longer such when 
the decomposition is performed in redder bands, or in the stellar mass 
profile. On the contrary, the disc scalelengths 
do not differ significantly in the different bands or going from the light
distribution to the actual stellar mass one. Some of the simulated galaxies
with large B/D ratios correspondingly display long scalelengths for the
remaining (quite small) disc component; but a proper comparison to observations
should consider only simulations with realistic B/D ratios for late--type
galaxies. 
In Fig.~\ref{fig:scalelengths} the solid lines are least square fits to the 
scalelength---circular velocity relation of simulated galaxies, limited 
to galaxies with {\mbox{(B/D)$_I \leq 0.6$}} (corresponding to 
{\mbox{(B/D)$_B \lsim 0.4$)}}\footnote{Notice that in recent deep surveys, 
the looser criterion {\mbox{(B/D)$<1$}}
is often used to distinguish late type galaxies from early types (e.g.\
Simard \etal 1999; Trujllo \& Aguerri 2004). The solid line fit in 
Fig.~\ref{fig:scalelengths} and our corresponding conclusions do not 
significantly change if simulated galaxies with (B/D) as large as $\sim$~1 
are included in the fit.}. 
In the I band (top right panel), 
this is compared to the observed relation derived by Sommer--Larsen \& Dolgov 
(2001) from the Mathewson sample ({\it dashed line}).
Our scalelengths are close to the observed average at low masses and
within a factor of 1.5 from the empirical relation for Milky Way--type 
galaxies ($V_c \sim 200$~km/sec); this is an improvement 
also on the results of SGP03 (their Fig. 8), since only for the largest 
circular speeds ($V_c \sim 300$~km/sec) we now find a factor--of--2 offset 
from observed scalelengths.

To compare properly to observational TF studies, the circular velocity $V_c$ 
of the system is measured at 2.2 disc scalelengths, by summing in quadrature
the contribution of the exponential disc component and of the (spherically
distributed) bulge and dark matter halo. As the angular momentum problem 
is not yet fully resolved (Fig.~\ref{fig:scalelengths}), the disc contribution
is computed by ``correcting'' its actual scalelength to the one expected
from the observed scalelength---$V_c$ relation (dashed line in 
Fig.~\ref{fig:scalelengths}). Namely we 
``redistribute'' the mass of the disc component according to the 
empirically expected scalelength for that circular velocity. 
$V_c$ is estimated at 2.2 
scalelengths and both quantities are adjusted to each other by iteration.
The procedure is as outlined in Sommer--Larsen \& Dolgov (2001) and the
main purpose is to avoid overestimating $V_c$ by measuring
it at too inner radii (due to the short scalelengths) where the rotation
curve of simulated galaxies typically presents a significant peak due to 
strong matter concentration in the central regions. 
Adjusting the scalelength to the ``observationally expected'' one we make sure
to measure $V_{2.2}$ far out enough that we skip the inner peak of the
rotation
curve and select regions where the circular velocity is only smoothly varying 
and representative of the global gravitational potential of the system. 
In fact, once the disc scalelength is adjusted around the actual observed
values, the corresponding $V_{2.2}$ is quite robust to e.g.\ the B/D ratio, 
to whether the bulge is considered point--like or modelled with an extended 
spherical symmetry, and so on. This shows clearly in Fig.~\ref{fig:barfrac},
where the dashed heavy line is the actual cumulative baryonic mass fraction
$(M_*/M_{tot}) (<R)$ in a simulated Milky Way--size galaxy, and the fainter
lines
are the estimated profiles of the baryonic contribution to the circular 
speed $(V_{2.2,bar}/V_c)^2$ after the disc mass has been ``redistributed''
as described above. The two fainter lines correspond to two different 
assumptions 
for the B/D ratio (solid: B/D=0.5; dashed: B/D=0); clearly when considering 
large enough galactocentric distances, the baryonic rotation speed is quite
robust to the detailed decomposition. For $V_c \sim$220~km/sec the empirical
scalelength is about 3.5~kpc (Fig.~\ref{fig:scalelengths}) so that in this 
example the circular velocity is estimated at about 8~kpc, far out enough
to be unaffected by decomposition. 

Notice that the adjustment of the scalelengths is not applied within 
the simulations --- where altering disc matter distribution would influence 
the star formation history etc. --- but only analytically {\it a posteriori}
and only for the sake of deriving the representative circular velocity of the
system; in Fig.~\ref{fig:scalelengths} and anywhere else in the paper we
consider the actual scalelengths resulting from the bulge--to--disc 
decomposition discussed above.

\section{The Tully--Fisher relation at \zeqzero}
\label{sect:TFz0}
In this section we compare our simulations to the observed local TF relation,
discuss the systematic offset between the two, and comment on resolution 
and numerical effects.

\begin{figure*}
\leavevmode
\psfig{file=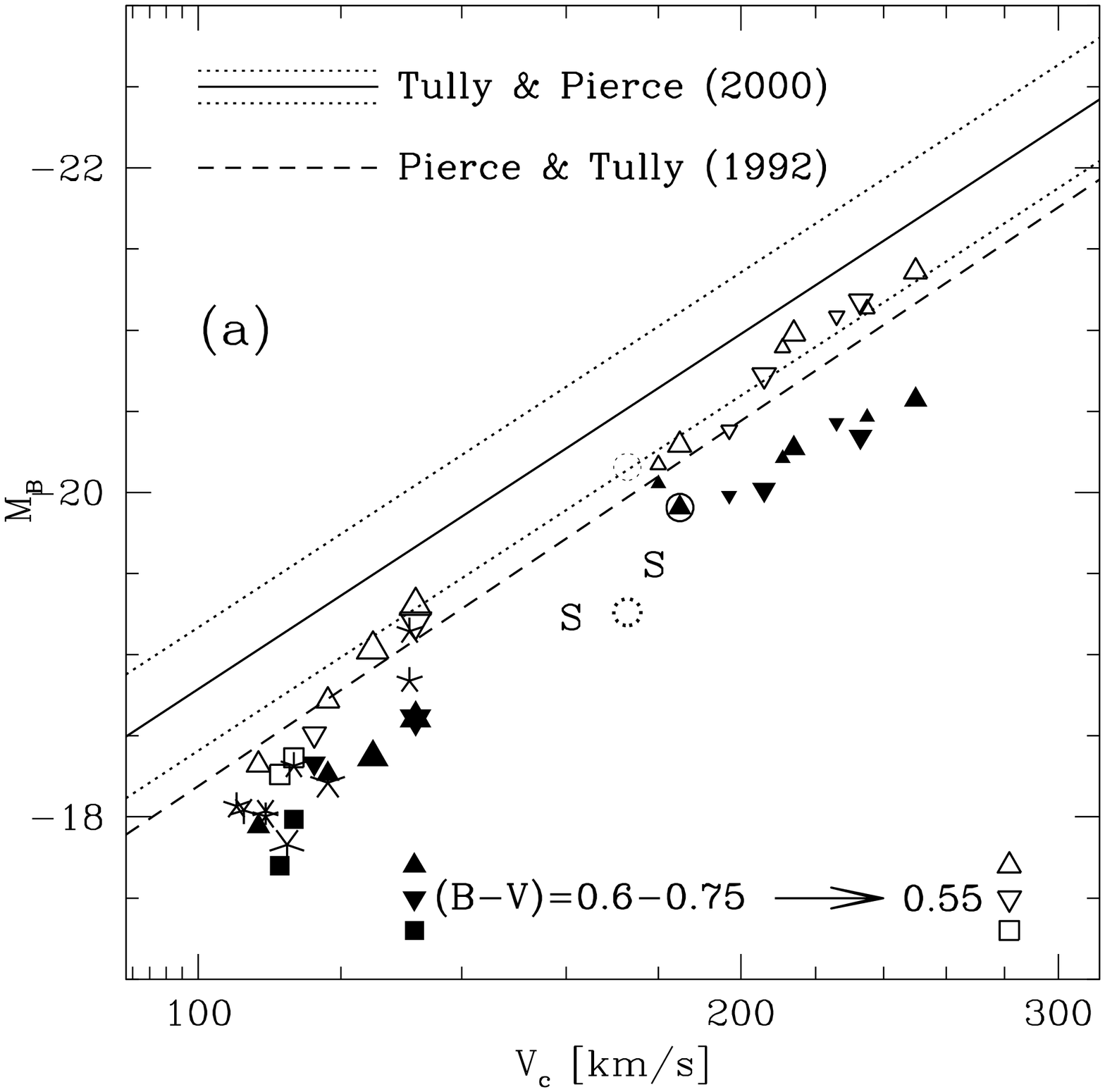,width=0.49\textwidth}
\psfig{file=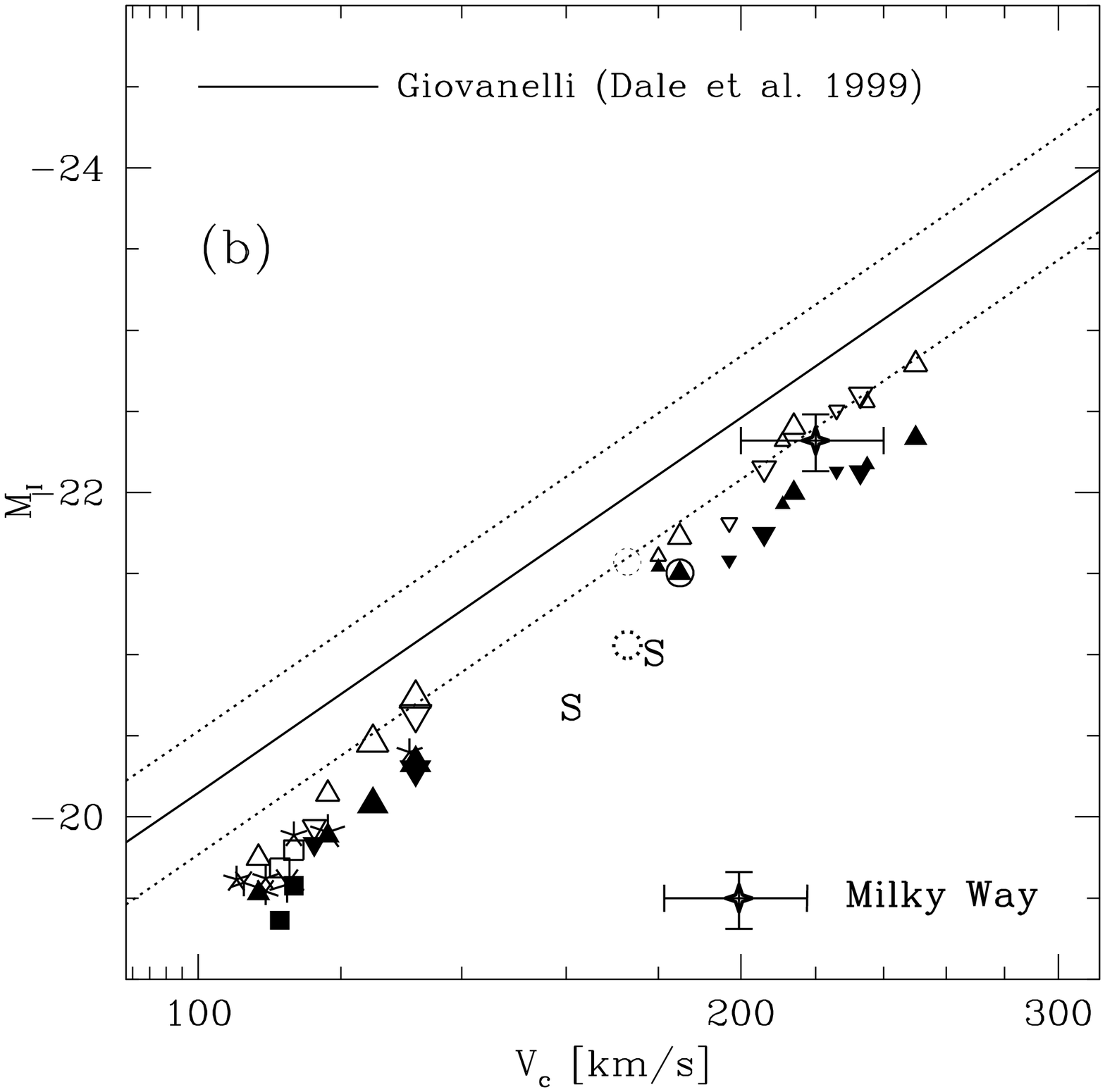,width=0.49\textwidth}

\leavevmode
\psfig{file=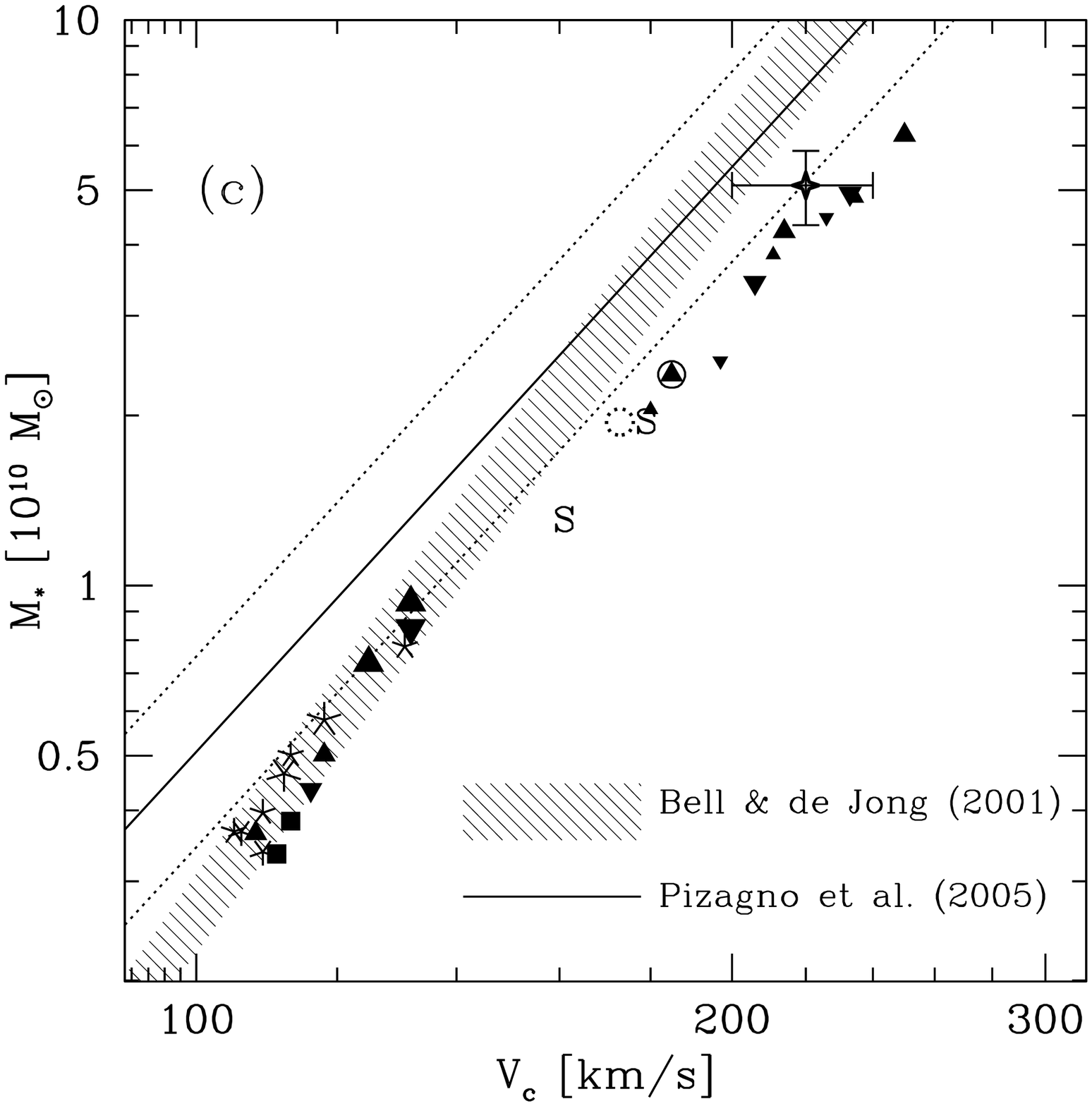,width=0.49\textwidth}
\psfig{file=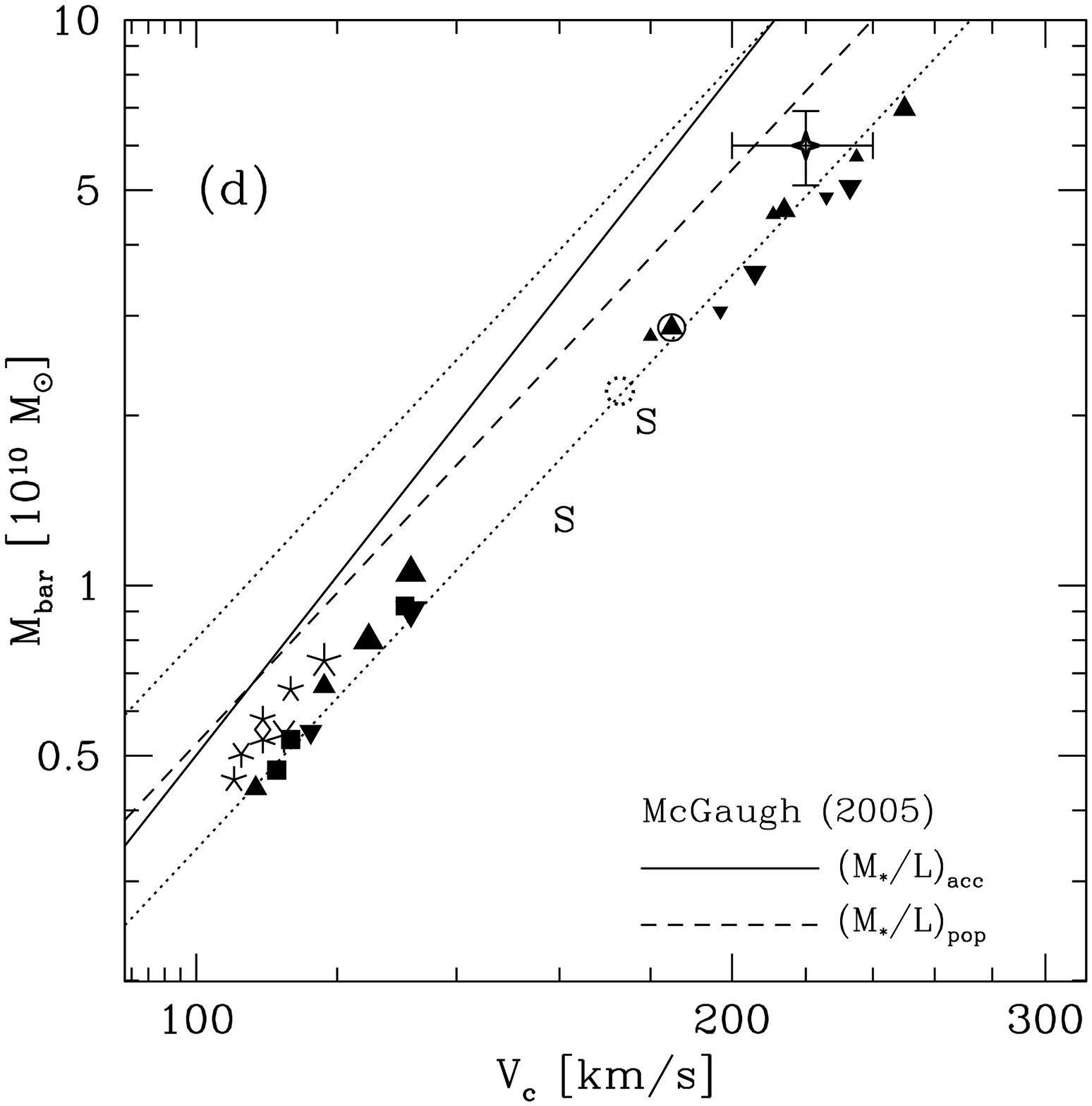,width=0.49\textwidth}
\caption{{\bf (a)} Simulated galaxies compared to the local ($z=0$) observed
Tully-Fisher relation in the B band and {\bf (b)} in the I band. 
{\it Solid dots}: direct simulation results; {\it open dots} ({\it cyan dots}
in the colour version of the figure): simulated
galaxies colour--corrected to B--V=0.55, more closely comparable to
the observational TF samples. Symbols are otherwise as in 
Fig.~\protect{\ref{fig:B/D}}. The size of the dots increases with the 
numerical resolution of the corresponding simulation; ``pairs'' of neighbouring
triangles (with the same orientation, pointing upward or downward, and same 
colour in the colour version of the figure) mark simulations of the same
object (or cosmological halo) resimulated at
two different resolution levels. Asterisks show simulations 
run in the ``long softening'' (\longsoft) mode. 
{\it Dotted circles}: the special ``no late
infall'' experiment discussed in Section~\protect{\ref{sect:noinfall}}; the
thin dotted circle is colour--corrected to B--V=0.55; the solid circle 
encircles the corresponding ``normal infall'' simulation.
The S symbols represent two galaxies simulated, at lower resolution, 
with the Salpeter IMF; only their colour--corrected location is shown
for simplicity, to be compared to the open symbols of the Kroupa IMF 
simulations.
{\bf (c)} Simulated galaxies compared to the empirical stellar
mass TF relation and {\bf (d)} baryonic TF relation (for two 
assumptions for assigning \ML; see text and McGaugh 2005 for details).
In panels b, c and d the data point with errorbars marks the location of the
Milky Way (from Flynn \etal 2006; see also Sommer--Larsen \& Dolgov 2001).}
\label{fig:TFz0}
\end{figure*}

\begin{figure}
\psfig{file=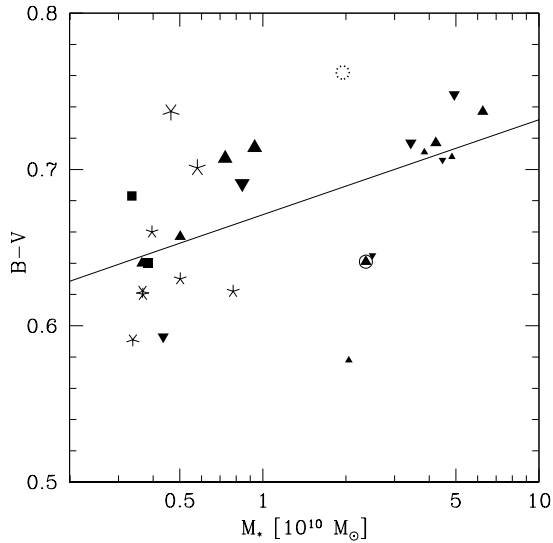,width=0.42\textwidth}
\caption{$B-V$ colours of simulated galaxies. Symbols as in 
Fig.~\protect{\ref{fig:TFz0}}. 
The solid line is a linear fit (excluding the
special ``no late infall'' experiment, namely the dotted circles) showing
that simulations trace a mass--colour relation. ``Pairs'' of triangles of
different size mark
the same object (or cosmological halo) resimulated at different resolution 
levels; higher resolution (larger symbols) produces larger stellar masses
and redder colours.}
\label{fig:colours}
\end{figure}

In Fig.~\ref{fig:TFz0} our simulations are compared to the B--band and I--band
Tully--Fisher relation (panels $a$ and $b$), to the stellar mass TF relation
(panel $c$) and to the baryonic TF relation (panel $d$). In the B band,
along with the most recent calibration by Tully \& Pierce (2000), we show
the older and fainter Pierce \& Tully (1992) calibration as this is still
widely used as a local reference for TF evolution studies; for both relations,
we adopted $W_R = 1.04 \times 2 \, V_c$ (Tully \& Fouqu\'e 1985).
Fig.~\ref{fig:TFz0}ab shows that the slope of the TF relation is well
reproduced, but our simulated galaxies (solid dots) are underluminous
with respect to the observed TF relation. Part of the offset can be explained
with differences in Hubble type: the observed TF relation is defined 
on samples of late type, Sbc--Sc spirals with a typical colour of $B-V$=0.55
(Roberts \& Haynes 1994). Our simulated galaxies (solid symbols) are of
somewhat earlier type
and redder colours ($B-V$=0.6--0.75, Fig.~\ref{fig:colours}); redder colours 
imply
larger stellar mass-to--light ratios (\ML) and induce systematic offsets
from the Tully--Fisher relation (Kannappan, Fabricant \& Franx 2002; PST04). 
The open symbols in Fig.~\ref{fig:TFz0}ab show in fact the colour--corrected
locus of the simulated galaxies, whose luminosities have been corrected 
to $B-V$=0.55 via the colour--\ML\ relations of PST04
(which agree well with the observed colour--dependent offsets of Kannappan 
\etal 2002).
Colour differences account for only about half of the offset in B and I band;
although, when the Pierce \& Tully (1992) calibration is considered,
there is good agreement in the zero--point after the colour corrections
are implemented.

The offset in Fig.~\ref{fig:TFz0}a would increase if the B band 
TF relations by Pierce \& Tully (1992), Tully \& Pierce (2000) were further
corrected by 0.27~mag of intrinsic face--on extinction (Tully \& Fouqu\'e 
1985), as is common in recent high redshift studies (e.g.\ B\"ohm \& Ziegler
2004; Bamford \etal 2006). In principle this should be a better comparison 
since our photometric model does not include dust; however, our colour 
correction to $B-V$=0.55 (open symbols) relies on the typical average colour 
of late--type spirals estimated from the RC3 catalogue (de Vaucouleurs \etal 
1991) corrected for inclination to face--on only, hence the comparison 
to face--on magnitudes is fair.

The problem of colour offsets is bypassed when multi--band data can be used
to infer the underlying stellar mass in the galaxies and define directly
the relation between circular velocity and stellar mass, i.e.\ the stellar
mass TF relation --- or the baryonic TF relation when the gas mass is also
included. 
Fig.~\ref{fig:TFz0}cd compares our simulated galaxies to the
stellar mass and baryonic TF relations, respectively; such comparisons 
are meaningful provided the same IMF is assumed in the simulations 
and in the empirical derivation of the stellar masses from multi--band data, 
which is indeed the case, to a first approximation.\footnote{Pizagno
\etal\ (2005) derived stellar masses from the colour--\ML\ 
relations of Bell \etal (2003), further scaled by --0.15~dex. Their relation 
is shallower and has 
a lighter normalization than the Kroupa IMF relations of PST04 (which apply
also to our simulated galaxies, and whose normalization has been confirmed 
in the recent study of the local Galactic disc by Flynn \etal\ 2006) or than
the ``scaled Salpeter'' relations of Bell \& de Jong (2001). 
However we have verified that, in the colour range of simulated galaxies, 
the (g-r)--\MLi\ relation adopted by Piz05 agrees to within about 0.05~dex
with the Kroupa IMF relation of PST04, so that their empirical stellar masses
can be consistently compared to our simulations (Fig.~\ref{fig:TFz0}c).

As to the baryonic TF relation, the $\cal P$=1 scaling factor adopted 
by McGaugh (2005) in his standard (\ML)$_{pop}$ TF relation closely 
corresponds to the Kroupa IMF 
normalization and to the photometric models adopted in our simulations, 
so that direct comparison is justified in Fig.~\ref{fig:TFz0}d.
Incidentally, we notice that with a scaling factor $\cal P$=0.7, 
the baryonic TF relation would shift into excellent agreement 
with our simulations, {\it both} 
in slope {\it and} in zero--point (cf.\ Table~2 of McGaugh 2005, for 
population synthesis scaling with $\cal P$=0.6--0.8). However, a scaling
to $\cal P$=0.7 corresponds to assuming that the IMF in disc galaxies 
is on average significantly lighter than the local Kroupa IMF 
(see also \S \ref{sect:differentIMF}).}

The slope of our stellar mass TF relation is in reasonable agreement 
with the recent results of Pizagno \etal (2005; henceforth, Piz05);
the relation derived earlier by 
Bell \& de Jong (2001) is significantly steeper, apparently a consequence 
of the different sample (cluster vs.\ field spirals, see the discussion 
in Piz05). In fact, the Bell \& de Jong
relation is based on the multi--band TF relation of Verheijen (2001) for
the Ursa Major cluster, which is steeper than e.g.\ the TF relations shown
in Fig.~\ref{fig:TFz0}ab; apparently the cluster environment affects the
properties of small spirals, steepening the TF relation with respect 
to field samples. Though the slope of the simulated stellar mass TF relation
well compares to that for field samples, Fig.~\ref{fig:TFz0}c shows again
a 1~$\sigma$ offset similar to that seen in panels a,b for the luminous 
TF relation.

In Fig.~\ref{fig:TFz0}d we plot the baryonic mass of our simulated galaxies,
summing the stellar mass and the cold gas mass, versus circular velocity
and compare their locus to the baryonic TF relation derived by McGaugh (2005).
The slope of the empirical baryonic TF relation is sensitive to how the stellar
\ML\ is assigned; the recipe favoured by McGaugh (2005), i.e.\ a \ML\ 
minimizing the scatter in the empirical mass discrepancy --- acceleration 
relation (and, consequently, in the TF relation; 
\ML$_{acc}$, solid line), implies a TF relation with a slope as steep as 4 
which is difficult to reproduce in the current hierarchical cosmological 
scenario. Conversely, relying on the predictions of stellar population 
synthesis models yields somewhat lower \ML's and a shallower TF relation 
(\ML$_{pop}$, dashed line),
with a slope closer to 3 which is the typical prediction of current cosmology,
and is in fact well reproduced by our simulations. Even in this case,
though, an offset of about 1~$\sigma$ is again found between simulated 
and empirical TF relation. (For clarity, 
in Fig.~\ref{fig:TFz0}d we show the 1~$\sigma$ scatter lines only for the 
(\ML)$_{pop}$ case; that of the (\ML)$_{acc}$ TF relation is about twice 
as small.)

\subsection{Resolution and numerical effects}
In Fig.~\ref{fig:TFz0}, the size of the symbols increases with the numerical 
resolution of the corresponding simulation, and ``pairs'' of neighbouring
triangles with the same orientation (pointing upward or downward)
are used to mark pairs of simulations of the same object (same cosmological 
halo) 
resimulated at different levels of resolution. Clearly, resolution has no 
systematic 
effects on the TF locus defined by simulated galaxies, although resimulations
at higher resolution tend to have systematically earlier star formation 
histories and redder colours (Fig.~\ref{fig:colours}), as expected since denser
regions are better resolved.

For $V_c < 150$~km/s a few simulations, marked by asterisks, are run 
in the ``long softening'' mode (\longsoft; see Section~\ref{sect:simulations}).
The \longsoft\ galaxies tend to be less concentrated and less 
bulge--dominated than their ``normal softening'' (\norsoft) counterparts, 
as expected; hence their structure and appearance resembles more closely 
that of real disc galaxies. However, no systematic differences are found 
in the location of the \longsoft\ vs.\ \norsoft\ galaxies in the TF plot
(Fig.~\ref{fig:TFz0}ab); the colour correction (open symbols in panels a, b)
is shown only for the
\norsoft\ simulations for clarity; but the effect on the \longsoft\
cases is similar.

We have also verified that these conclusions, i.e.\ that numerical resolution 
and choice of softening length induce no systematic biases on the TF locus, 
hold also at higher redshifts. When discussing the evolution of the TF relation
in Section~\ref{sect:TFevolution}, we will mainly consider and plot, 
for $V_c < 150$~km/s, the \norsoft\ simulations;
but we stress that, for the sake of the TF locus at any redshift, 
no systematic difference in the results is introduced if the \longsoft\ 
simulations are considered istead.

\subsection{Interpreting the offset}
The offset between simulations and observations seen in all panels 
of Fig.~\ref{fig:TFz0} is typically imputed to an excess of dark matter 
in the central regions of simulated galaxies (Navarro \& Steinmetz 2000ab).
In fact, circular velocity
traces the total gravitational potential of a galaxy, i.e. its total mass; 
if, at a given luminosity or baryonic mass, simulated galaxies 
are rotating faster than observed, this means that their total (dark+luminous) 
mass is larger than it should empirically be, indicating an excess of dark
matter. In the language of rotation curve studies,
this corresponds to the well known problem of the too large
concentration of dark matter haloes predicted by CDM.

\begin{figure}
\psfig{file=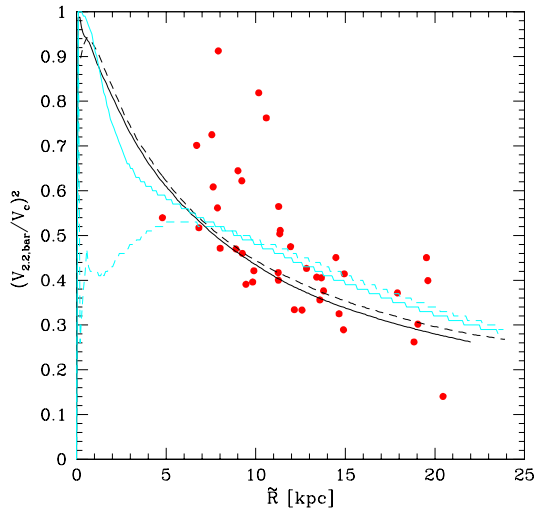,width=0.46\textwidth}
\caption{Fractional contribution of the baryons to the total circular
velocity, as a function of the ``scaled'' radius 
$\tilde{R}= R \left( \frac{220}{V_c} \right)$, with $R=2.2$~scalelengths
(see text). 
{\it Lines:} simulated galaxies; {\it dots:} galaxy sample of Pizagno \etal
(2005).}
\label{fig:barfrac}
\end{figure}

\subsubsection{A central ``cusp'' problem?}
We investigate the dark matter excess in our simulated galaxies as follows.
Figure~\ref{fig:barfrac} shows the contribution of the baryons to the total
circular velocity for the sample galaxies of Piz05 with $V_c>$150~km/sec; 
these are plotted as a function of the ``scaled'' radius 
$\tilde{R}= R \left( \frac{220}{V_c} \right)$ (so as to scale all galaxies 
to approximately the same linear extent), where $R$ equals 2.2 
scalelengths. The heavy lines show the mass 
ratio profile $(M_*/M_{tot}) (<R)$, also scaled in terms of 
$\tilde{R}$, for two simulated Milky Way--type galaxies 
($V_c \sim 230-250$~km/sec). The faint lines display, for one of these two
galaxies, the $(V_{2.2,bar}/V_c)^2$ profile we estimate by ``redistributing''
the disc mass according to its empirically expected scalelength, as discussed
in Section~\ref{sect:analysis} for the determination of the circular speed.
The two faint lines correspond to two different assumptions for 
the B/D ratio (0.5 for the solid line, 0 for the dashed line).

Three cases, where the stellar mass inferred 
by Piz05 is actually larger than allowed by the rotational speed, 
i.e. $(V_{2.2,bar}/V_c)^2 > 1$, have been excluded. We recall here that
a ``maximal disc'' typically contributes to 85\% of the rotation curve at
2.2~scalelengths (Sackett 1997), or $(V_{2.2,bar}/V_c)^2 \sim 0.7$. 
A few of the galaxies in the Piz05 sample seem thus to be 
``supermaximal''. Excluding such objects from the discussion, it seems 
that our simulations trace the cumulative baryonic mass fraction reasonably
well, once we consider radii sufficiently far out.

As detailed in Section~\ref{sect:analysis}, we compute our $V_{2.2}$ adopting
some ``empirically corrected'' scalelength and disc mass distribution, 
to avoid the known problem of too small and concentrated discs, 
which could bias our $V_c$ measurement 
in the simulations. This should limit the ``cusp'' influence on our 
$V_c$ estimate, and Fig.~\ref{fig:barfrac} shows that this is indeed
the case for large enough radii. For a Milky Way--type galaxy 
($V_c \sim$220~km/sec, $h_I \sim$3.5~kpc) we estimate $V_c$ at about 8~kpc,
where indeed the baryon/DM proportion in the rotation curve looks realistic.

\subsubsection{A bias to larger scalelengths?}
Possibly contributing to the TF discrepancy in Fig.~\ref{fig:TFz0}c is 
the fact that the SDSS spirals of the Piz05 sample have significantly larger 
scalelengths than the average relation holding for the larger and more
representative Mathewson sample, which we have 
adopted for our scalelength correction (Fig.~\ref{fig:scalelengths}). 
That the Piz05 sample is biased to large scalelengths, 
possibly as a results of their extreme sample selection (B/D)$_i \leq 0.1$,
is also highlighted by Dutton \etal (2006).
We have checked whether adopting the Piz05
$V_c$--scalelength relation to select (further out than before) the radius
$R=2.2$ scalelengths where to measure our $V_{2.2}$,
would reduce the offset in Fig.~\ref{fig:TFz0}c;
but the effect on the estimated $V_c$ is less than half the offset. 
So, larger scalelengths are not the full solution
and the cause of the TF offsets must lie elsehwere.

\subsubsection{A different Initial Mass Function?}
\label{sect:differentIMF}
One possible culprit is the adopted stellar IMF: the Kroupa IMF 
is a ``bottom--light'' IMF yet it is still at least 10\% too ``heavy'' 
than required to reproduce the zero--point of the I band TF relation 
(see e.g.\ Fig.~4 in PST04); however a further 10\% or 0.1~mag offset 
still would not reconcile 
simulations and observations in Fig.~\ref{fig:TFz0}b. It is noteworthy 
though in this respect, that the Milky Way itself, whose chemical 
and photometric properties are well described by a Kroupa IMF, 
is also underluminous by about 1~$\sigma$ with respect to the observed 
TF relation (Flynn \etal 2006; see
the data point with errorbars in Fig.~\ref{fig:TFz0}), this is 
about the same offset as our simulations.
Indeed the Milky Way lies much closer to the TF locus of the simulated
galaxies, than to the empirical TF relation --- namely, our simulations
with the Kroupa IMF nicely match the Milky Way but not the TF normalization.
This may suggest that the typical average IMF of external disc galaxies 
is more ``bottom--light'' than the local Solar Neighbourhood one 
(see also Gnedin \etal 2006). 

The effect of changing the IMF (to a more bottom--heavy one) is shown 
by the two lower resolution runs simulated with the Salpeter IMF, represented 
by the S symbols in Fig.~\ref{fig:TFz0}; in panels {\it a} and {\it b}, only 
the colour--corrected location of these two galaxies is shown for simplicity,
to be compared to the open symbols for the Kroupa IMF galaxies.
The $M_*/L$ ratio of the Salpeter IMF is higher, therefore the S
galaxies define a significantly dimmer zero--point for the B and I-band TF 
relation. However, the stellar (or baryonic) mass TF normalization
is hardly affected by the change in the IMF
(and by the related change in supernova rates, feedback and chemical enrichment
efficiency, gas restitution rate etc.):
the baryonic mass that cools out
to form the stellar disc correlates with the resulting
circular velocity, so that data points in the ($M_*, V_c$) plane tend 
to move along the TF locus leaving the zero--point unaffected 
(Navarro \& Steinmetz 2000a,b).

Likewise, adopting a more bottom--light IMF than the Kroupa one
in the simulations would also hardly affect the stellar mass TF locus.
However, the $M_*/L$ ratio would be lower hence the TF locus would be brighter 
in the ($L, V_c$) plane. Besides, the {\it observed} stellar/baryonic mass 
TF relation 
in Fig.~\ref{fig:TFz0}c,d would be lighter if a more bottom light IMF 
were assumed to translate multi--band photometry to stellar mass.

\subsubsection{A different Universe?}
\label{sect:diffUniv}
A little bias in $M_*/L$ ratio is also expected from our choice of 
$h=0.65$, which makes our simulated Universe and galaxies $\sim$1~Gyr
older than with the presently more popular choice of $h=0.7$. Systematically 
larger galaxy ages indeed correspond to systematically larger $M_*/L$ ratios,
even for the same colour; but the effect over just 1~Gyr age offset is expected
to be minor (Bell \& de Jong 2001). One of the two Salpeter galaxies 
in Fig.~\ref{fig:TFz0} is actually run in a $h=0.7$, $\sigma_8=0.9$ 
Universe; but no significant difference in its TF location is seen,
with respect to the other Salpeter galaxy; hence the effect of the
younger Universe is minor.

The zero--points of simulated TF relations also depends on the cosmological
parameters, especially $\Omega_0$ and $\sigma_8$, via the baryon fraction
and the concentration of the dark matter haloes
(Avila--Reese, Firmani \& Hern\'andez 1998; van den Bosch 2000;
Navarro \& Steinmetz 2000a,b; Eke, Navarro \& Steinmetz 2001;
Buchalter, Jimenez \& Kamionkowski 2001).
Our simulations are run in the ``concordance'' $\Lambda$CDM model
--- and as mentioned in the previous paragraph, choosing $\sigma_8=1.0$
over 0.9 does not make a significant difference;
but the present ``concordance'' on the exact cosmological parameters 
is not perfect and values as low as $\Omega_M=0.26$ and especially 
$\sigma_8=0.74$ are advocated by the recent 3--year WMAP data release
(Spergel \etal 2006; see also van den Bosch, Mo \& Yang 2003). This revised
cosmology results in lower halo concentrations, aiding the match with the
zero--point of the TF relation (Gnedin \etal 2006), although some 
further effects counteracting
adiabatic contraction may still need to be invoked (Dutton \etal 2006).

\subsubsection{Adiabatic contraction?}
\label{sect:pinching}
We have also examined the role of adiabatic contraction with some additional
experiment on our most massive simulated galaxy (at $V_c \sim 250$~km/sec).
An excess in adiabatic contraction of the dark halo may occur in our 
simulations following the high baryonic concentration, as the angular momentum
problem is not fully resolved, especially for massive galaxies.
Although, when we compute $V_c$, we correct for the remaining angular momentum 
problem by ``redistributing'' the disc mass according to the empirically 
expected scalelength, we do not usually correct for the exceeding dark matter 
``pinching'' that has occurred. The resulting dark matter content of our 
simulated galaxies within 2.2 scalelengths (the radius where we measure $V_c$) 
is of the order of 50\%, ranging from 0.4--0.45 for our most massive objects
($V_c \sim$200--250~km-sec) to 0.55--0.6 for the dwarfs (100--120 km/sec).

For our most massive galaxy, we estimate
the significance of the over--pinching by slowly (adiabatically) removing 
the present stellar disk+bulge, and inserting instead a pure exponential disk 
of mass $6.9 \times 10^{10}$~\Msol, and scalelength $R_d=3.8$~kpc 
(cf.\ SGP03). The resulting $V_c$ is reduced by less than 3\%. Even neglecting
adiabatic contraction completely, by combining the expected baryonic disc 
profile described above with the dark halo profile resulting from a
DM--only simulation of the same object, reduces the resulting $V_c$ 
by only 6\% or so.
The effects of adiabatic contraction are by far not enough to cure the TF 
offsets in Fig.~\ref{fig:TFz0}; in fact adiabatic {\it expansion} has been 
recently invoked in the literature to match the TF zero--point 
(Dutton \etal 2006).

\bigskip
Possibly a combination of all the contributing effects discussed above 
will help reproducing the zero--point
of the TF relation; however, we remark that some uncertainty may still exist 
in the empirical zero--point itself: as mentioned above, the Milky Way 
for instance is also underluminous with respect to it (Flynn \etal 2006;
see Fig.~\ref{fig:TFz0}b).
Also, for some of the existing normalizations, e.g.\ the Pierce \& Tully 
(1992) TF relation which is still widely used as a local reference for TF
evolution studies, no significant discrepancy remains 
once the colour correction is accounted for 
(Fig.~\ref{fig:TFz0}a).

\begin{figure*}
\leavevmode
\psfig{file=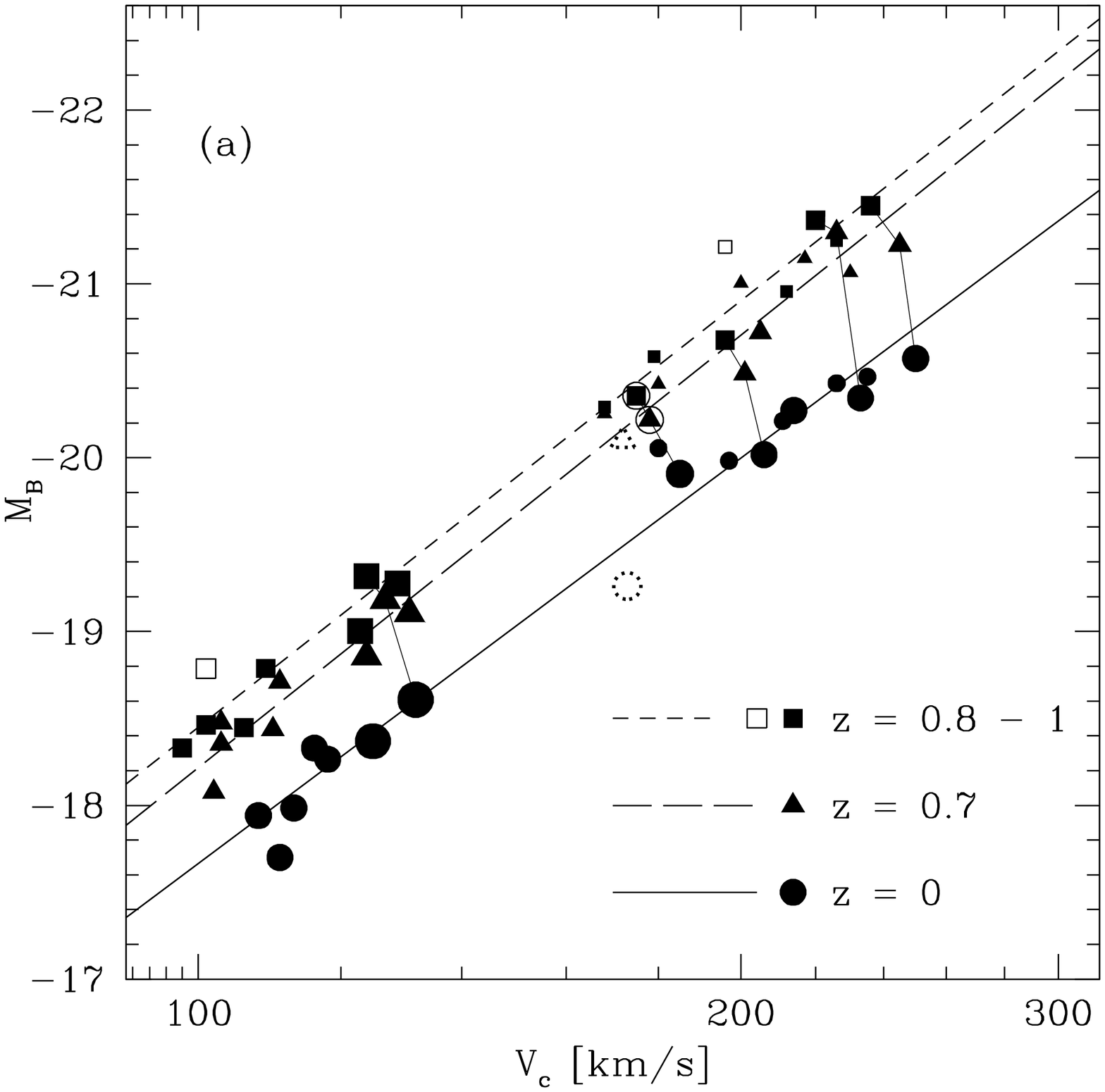,width=0.49\textwidth}
\psfig{file=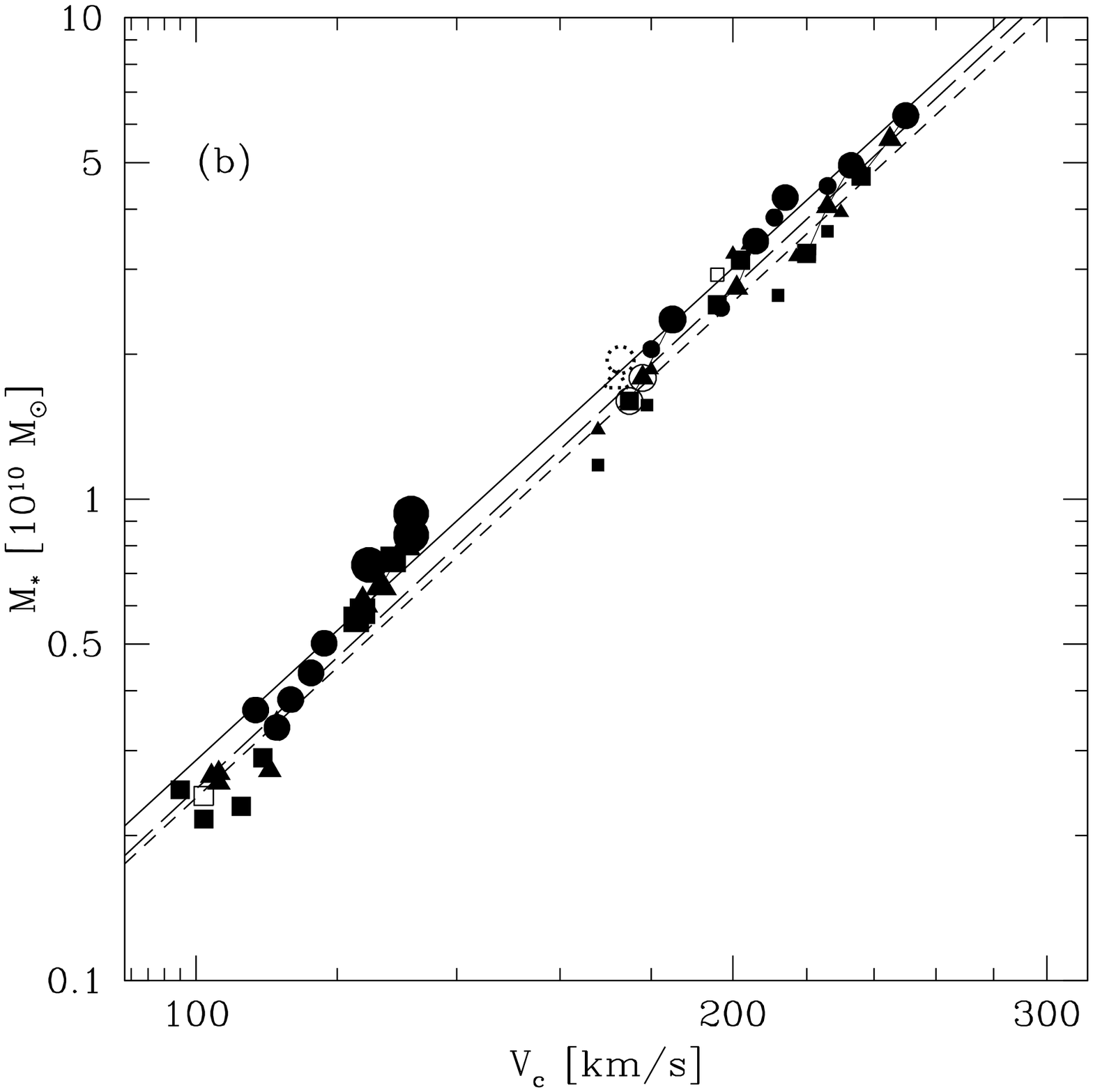,width=0.49\textwidth}
\caption{{\bf (a)} B band and {\bf (b)} stellar mass TF relation of simulated 
galaxies at $z=0$ (circles), $z=0.7$ (triangles) and $z=1$ (filled squares;
open squares trace a few objects analyzed at $z=0.8$ instead of $z=1$). For
a few objects, the evolution from $z=1$ to 0 is connected by a thin line for
the sake of example. The size of the dots increases with the numerical 
resolution of the corresponding simulation. For $V_c < 150$~km/s, only
``long softening'' (\longsoft) simulations are shown for clarity. The solid,
long--dashed and short--dashed lines are least square fits to the TF relation
at $z=0$, $z=0.7$ and $z=1$ respectively.}
\label{fig:TFevol}
\end{figure*}

\section{The evolution of the Tully--Fisher relation}
\label{sect:TFevolution}

In this section we analyze the evolution of the TF relation as predicted
by our simulations, both in terms of luminosity and of stellar mass.

In Fig.~\ref{fig:TFevol}a we plot the TF relation in the B band for our 
simulated
galaxies at different redshifts: $z=0$ (full circles), $z=0.7$ (triangles) and
$z=1$ (full squares). Two of the simulated galaxies are quite disturbed at
$z=1$ and were analyzed at $z=0.8$ instead (open squares).
The solid, long dashed and short dashed lines are least square fits to the
data points at $z=0$, 0.7 and~1 respectively. (The dotted symbols,
representing the special ``no late infall'' experiment discussed 
in Section~\ref{sect:noinfall}, are excluded from the fit). For circular
velocities $V_c < 150$~km/s, only \norsoft\ mode simulations are shown 
in Fig.~\ref{fig:TFevol} and included in the various fits. This is just for the
sake of clarity in the figure, as we verified that including the \longsoft\
simulations does not change the overall picture nor the fits significantly.
It is clear from Fig.~\ref{fig:TFevol}a that the TF relation is predicted
to evolve in luminosity, maintaining the slope nearly constant ---
notice in fact that that we did \underline{not} impose a constant slope
in the formal fits of Fig.~\ref{fig:TFevol}a. The slope
is around 8, i.e.\ close to $x=3$ when expressed as $L \propto V_c^x$, 
as expected in the current cosmological scenario (Navarro \& Steinmetz 2000; 
Sommer-Larsen 2006).\footnote{Formally the fits yield a marginal 
steepening of the slope, from 7.75 at $z=0$ to about 8.2 at $z$=0.7--1; 
this trend is in the same direction, but far milder (hardly significant) 
than the recent findings of Weiner \etal (2006) from their large TKRS/GOODS 
galaxy sample. This is related, as discussed by Weiner \etal, to more massive
galaxies being tendentially redder (Fig.~\ref{fig:colours}) hence passively
fading in luminosity at a more rapid pace between $z=1$ and~0. 
Our stellar mass TF relation, on the other hand, shows a fairly constant
slope of 3.4 with no steepening at all.}
The evolutionary offset with respect to the $z=0$ relation is 
$\Delta M_B \sim 0.7$~mag at $z=0.7$ and $\Delta M_B \sim 0.85$~mag at $z=1$.
This is somewhat intermediate in the range of current literature results
(see the Table~\ref{tab:literature} in the introduction), but is certainly
not compatible with very mild luminosity evolution.

The two galaxies analyzed at $z=0.8$ (because too disturbed at $z=1$;
open squares in Fig.~\ref{fig:TFevol}) appear to be slightly overluminous
with respect to other ``quiet'' objects at $z=1$; this suggests that recent
disturbance or dynamical interactions may enhance the luminosity with
respect to the ``normal'' TF relation, as found in the local TF samples
by Kannappan \& Barton (2004). We verified, however, that their inclusion 
in the $z=1$ fit has negligible effects, hence they are considered together
with all the other $z=1$ objects.

The luminosity evolution in Fig.~\ref{fig:TFevol}a is driven by aging and
dimming of the stellar populations hosted in the galaxies; while in terms
of stellar mass, the evolution of the TF relation up to $z=1$ is rather
negligible (Fig.~\ref{fig:TFevol}b). At a given circular velocity, the
increase in stellar mass between $z=1$ and $z=0$ is around 0.07--0.1~dex.

We stress, though, that the very mild evolution of the stellar mass applies
only to the TF as a relation, but not to the individual objects, which
typically grow by a factor of 1.5--2 in baryonic mass, 
between z=1 and z=0 (some examples of evolution of individual objects 
are connected by the thin lines in Fig.~\ref{fig:TFevol}).
As a consequence, interpreting the evolution of the TF relation 
as a pure luminosity evolution {\it applied to the individual galaxy}
at fixed $V_c$ (e.g.\ Weiner \etal 2006) may be misleading. In fact, 
although the circular velocity of the inner dark halo regions is expected 
to evolve negligibly at late epochs (Mo \etal 1998; Wechsler \etal 2002), 
the growth of the disc mass by infall of halo gas and on--going 
star formation is necessarily accompanied by an increased disc contribution 
to the measured total $V_c$ of the galaxy, which thus grows in time.
Because of the corresponding growth in stellar mass {\it and} circular
velocity, however, the evolution of a given galaxy largely occurs along 
the TF relation so that the relation {\it per se}, considered at fixed $V_c$, 
evolves much less than the individual objects. 

Recent observational results (Weiner \etal 2006) highlight indeed that 
the intercept of the TF relation evolves far more strongly in B band than
in the near--infrared (J band), which is a closer tracer of stellar mass
--- although they also find a significant evolution in the slope, which we
do not see in our simulated TF relation.

A dichotomy in the evolution of the TF relation observed in different bands 
was predicted also by e.g.\ Firmani \& Avila--Reese (1999) and
Buchalter \etal (2001). The latter results are qualitatively similar to ours,
with a K band TF relation (closer to the actual stellar mass TF) that is very
similar between $z=0$ and $z=1$, and a B band TF relation which is 
significantly brighter (and marginally steeper) at $z=1$. However 
the brightening they find is more significant than ours (larger than 1~mag);
this is likely due to the fact that in their simplified models the disc is
already in place at the redshift of formation, and after the initial peak
(when the gas mass is largest) its star formation history is ever declining;
hence their discs undergo purely passive evolution, while in our simulations
they gradually build up via mergers, accretion and infall of halo gas --- which
must partly counteract the trend of passive luminosity dimming.

Firmani \& Avila--Reese (1999), with their more detailed cosmological 
semi--numerical models, 
considered the TF relation in B band and in H band.
They predict as well that the slope of the TF relation remains
fixed with redshift, however they find that
the H band TF zero--point gets significantly fainter at increasing redshift,
in contrasts to our results of very minor evolution of the stellar mass
TF relation. In the B band, the evolution of their TF zero--point
reverses, but the brightening they find, out to $z \sim$1, is much milder 
than our results. The difference between our respective predictions apparently
resides in the fact that in their model galaxies, while the stellar mass
steadily increases, their $V_{max}$ remains roughly constant, rather than 
correlate with $M_*$ along the TF slope as we discussed above for our 
simulation.

The correlation we find between $V_c$ and $M_*$ may be artificially enhanced
by an excess in adiabatic contraction of the dark halo following
the high baryonic concentration of simulated galaxies. However, as we discussed
in Section~\ref{sect:pinching}, the estimated effect of such ``over--pinching''
is less than 3\%, therefore our $V_c$'s are likely not over--sensitive 
to the disc mass.
Conversely, in the models of Firmani \& Avila--Reese the baryonic disc 
possibly does not dominate enough the rotation curves and their $V_{max}$ 
does not correlate sufficiently with the disc mass; some correlation 
should indeed be expected since
the baryonic contribution to rotation curves out to 2.2 scalelengths
is prominent. In this sense, the evolution of the stellar mass 
TF relation can be considered as a probe of the baryon dominance and of the
dark matter/baryon correlation in the rotation curve of disc galaxies.

\section{A special ``no late infall'' experiment}
\label{sect:noinfall}
For one of our simulations we run a special experiment, artificially halting
infall of hot, dilute gas onto the disc at late times ($z<1$) by removing all 
low density gas (log($n_H$)$<-2$, log($T$)$>$30000~K) from the halo 
of the galaxy 
at $z=1$, leaving just two satellite galaxies in the halo.
 
This experiment is useful in view of the current debate on disc 
accretion: is a significant late infall of halo gas onto the disc compatible 
with the observed X--ray luminosity of galactic haloes (Toft \etal 2002; 
Rasmussen \etal 2004; Pedersen \etal 2005) and crucial for its build up, 
as typically assumed in standard galaxy formation scenarios? Or is disc galaxy
evolution at late times driven by other mechanisms, like accretion of 
satellites (e.g.\ Helmi \etal 1999, 2006) or of cold gas
(Birnboim \& Dekel 2003; Binney 2005)?
In this experiment, infall of hot and dilute halo gas is prevented, 
but satellite or cold gas accretion onto the galaxy can still occur.

Another useful application of this experiment is the evolution of disc
galaxies in clusters: cluster spirals observed at intermediate redshifts 
apparently transform into passive spirals and S0's by the present day 
(Dressler \etal 1997; Fasano \etal 2000; Smith \etal 2005), and quenching 
of star formation of the galaxies infalling onto the cluster is apparent
from their photometric and spectroscopic properties (the Butcher--Oemler 
effect: Butcher \& Oemler 1978, 1984, Ellingson \etal 2001, Margoniner 
\etal 2001, Kodama \& Bower 2001; galaxies with k+a spectra: Dressler 
\& Gunn 1983, Couch \& Sharples 1987, Poggianti \etal 1999, 2004; 
or with emission lines: Lewis \etal 2002, Gomez \etal 2003).
One of the possible culprits is the stripping of the halo gas reservoir when 
galaxies enter the cluster environment and the intra--cluster medium (ICM;
Dressler 2004 and references therein).
A crude way to simulate the ICM stripping is in fact to remove the hot,
low density halo
gas from the simulated galaxy.

In Fig.~\ref{fig:TFz0} this special experiment 
is indicated by the dotted
circle at $V_c \sim 175$~km/s --- and the corresponding ``normal'', 
i.e.\ non--stripped, galaxy is marked by the triangle inscribed within 
a circle at $V_c \sim 185$~km/s. The effect of halo gas stripping 
is to reduce the fuel for star formation, thus reducing the final stellar mass 
and luminosity of the galaxy, as well as its circular velocity; the net effect 
is to let it settle on the same stellar-baryonic mass TF relation, 
albeit at a lower $V_c$ (Fig.~\ref{fig:TFz0}cd). The stripped galaxy is
fainter than the TF relation defined by the other galaxies 
(Fig.~\ref{fig:TFz0}ab), but it is also significantly redder than its
``normal'' counterpart (Fig.~\ref{fig:colours}) so that, once the colour
correction is applied, the galaxy lies on the same TF relation as all the
others (thin dotted circle in Fig.~\ref{fig:TFz0}ab). Henceforth, with respect
to TF properties the stripped galaxy at $z=0$ looks like a standard
early--type spiral, fainter than the standard TF relation just due to
its redder colours. As a ``cluster spiral'' experiment, this indicates that
the TF relation for cluster galaxies at $z=0$ should not be markedly different 
from the field one of similar Hubble type, or once colour--corrections 
are applied. Notice though that we have only considered a relatively massive
object for this experiment, while cluster environments may affect more
importantly the TF relation at the low mass end (see Fig.~\ref{fig:TFz0}c and
the related discussion in Section~\ref{sect:TFz0}).

In redshift, the stripped galaxy evolves more significantly in magnitude 
than the ``normal''
TF relation (Fig.~\ref{fig:TFevol}a) and as an individual object its evolution 
occurs typically at constant $V_c$, as halo gas stripping has quenched 
its build--up 
by infall at $z>1$. This may hint to the possibility that, if halo gas infall
were indeed irrelevant at late times, the TF relation would evolve more
significantly than predicted by the standard simulations. Unfortunately,
due to the present observational uncertainty in the TF evolution, it is not
yet possible to distinguish between the various scenarios.

This experiment, artificially creating an early--type spiral, also highlights 
that the extent of the TF evolution likely depends on the final Hubble type 
of the spirals considered: the TF for early type galaxies is likely to evolve 
more significantly than that of late--type galaxies. When studying high 
redshift samples however, one hardly knows whether one is sampling 
the progenitors of Sa's or of Sc's, hence it may be misleading to compare
to standard literature Sbc-Sc TF samples at $z=0$ born chiefly as distance 
indicators (e.g.\ Pierce \& Tully 1992). The Hubble--type effect might 
contribute to the present wide range of results obtained by different authors
on the TF relation. As pointed out by Kannappan \& Barton (2004), one should 
rather compare to a volume--limited local sample selected with analogous 
criteria as the high--redshift one; not only to include in the comparison
the role of dynamical disturbances that may bias the high--$z$ samples 
bright (and lead to overestimate the TF evolution), but also to compare
locally to a fair mixture of Hubble types: if the local TF relation 
is preferentially based on the (brighter) Sbc-Sc types, the corresponding
TF evolution may be underestimated. The Hubble type effect is by no means
negligible: from Sc to Sb to Sa, TF offsets of 0.4 to 0.7 mag are found
(Kannappan \etal 2002; PST04), i.e.\ significant with respect to the observed
amount of TF evolution (0--1.5~mag, see the Introduction).

Clearly the evolution of the stellar mass TF relation bypasses the Hubble type
bias, however this is much more demanding observationally as multi--band
data are required.

\section{Summary and conclusions}
\label{sect:conclusions}
In this paper we discuss the TF relation and its evolution as predicted
by cosmological simulations of disc galaxy formation including hydrodynamics 
(Tree-SPH) and all the relevant baryonic physics, such as star formation,
chemical and photometric evolution, metal--dependent cooling, energy feedback.

The resulting disc galaxies at $z=0$ have scalelengths compatible with the
observed ones in the low--mass range ($V_c \sim$100--120~km/sec), and somewhat
shorter, yet within a factor of 2, for Milky Way--type or more massive objects.
Though still somewhat plagued by the angular momentum problem, 
our physical recipe and implementation of strong early feedback goes 
a long way in creating realistic galaxies (SGP03).

At $z=0$, the TF relation defined by our simulated galaxies is offset
by about 0.4~mag in luminosity, and by about 40\% in stellar 
(and baryonic) mass, with respect to observational TF relations. 
This is a well known problem of galaxy formation simulations
(Navarro \& Steinmetz 2000) and the origin of the offset remains somewhat 
unclear. We demonstrate that it is likely not due to the dark matter
cusp in the central regions, and that it cannot be cured by simply avoiding
adiabatic contraction. A different choice of cosmological parameters 
(such as $\sigma_8$) and/or 
of stellar IMF in disc galaxies may concur to solve the problem; some mechanism
for adiabatic expansion of the halo from the baryons has also been recently
invoked in this respect (Dutton \etal 2006).

However, we remark that the location of the Milky Way with respect 
to the TF relation is also offset (Flynn \etal 2006), by as much as our 
simulations. The fact that our Galaxy lies on the same TF locus 
as our simulations adopting a Solar Neighbourhood--like IMF 
(Fig.~\ref{fig:TFz0}), suggests the puzzling possibility that the
problem may not lie with the simulations or the cosmology, but rather
with the luminosity zero--point of the observed TF relation of external 
galaxies, or with the IMF of the Milky Way which may not be representative
for all disc galaxies.

As to the luminosity evolution of the TF relation with redshift, 
there is presently no consensus observationally, with estimates ranging 
from negligible evolution to $\Delta M_B >$1~mag by $z \sim$1. Also, there 
are claims that the evolution may be mostly in the slope (with a stronger
brightening for lower mass objects) rather than a systematic offset, but this 
also is still debated. There is however some preliminary consensus that
the stellar mass TF relation hardly evolves out to $z \sim$1.

Our simulations predict a significant B--band luminosity evolution
($\Delta M_B$=0.7~mag by $z$=0.7 and $\Delta M_B$=0.85~mag by $z$=1),
while the TF slope remains nearly constant (systematic zero--point evolution).
The redshift evolution of the TF relation is a result of the 
correlated increase in stellar mass and $V_c$ in the individual galaxies, 
combined with luminosity fading and reddening caused by the aging of the host 
stellar populations. At fixed $V_c$, the net effect is a systematic decrease
in B band luminosity, while in terms of stellar mass the TF relation hardly 
evolves at all, in agreement with the present available observational evidence.

This does not mean that the stellar mass content of the individual 
galaxies does not evolve: each galaxy grows typically by 50--100\% in mass 
between $z=1$ and $z=0$, but as its baryonic mass grows, its circular velocity 
also increases so that the evolution of the individual object occurs largely
{\it along} the TF relation. The relation as such does not show any 
significant evolution.

We also consider as a special experiment a simulation where late gas infall
on the disc is precluded, by artificially stripping the galaxy of its hot, 
dilute halo gas at $z=1$. This is useful to assess alternative scenarios 
of disc
galaxy formation, where late disc growth is driven by other processes
(like satellite or cold gas accretion), as well as a possible scenario
of spiral evolution in clusters, where ram pressure stripping from interaction
with the ICM is a possible driver of morphological transformations. We find
that, from the point of view of the TF properties, the resulting object is
indistinguishable from early type spirals with little recent star formation
and red colours --- henceforth the TF relation for cluster galaxies at $z=0$
should not be markedly different from the field one, once colour--corrections
are applied. Its TF evolution is however more significant than that of the
``normal'' simulations, suggesting that, if halo gas infall had no major role 
in disc galaxy evolution at late times, the TF relation would show a much 
stronger luminosity evolution. Unfortunately the present observational
uncertainty does not allow to discuss if either scenario is really favoured.
This experiment, artificially creating an early--type galaxy, also underlines
that the amount of TF evolution likely depends on Hubble spiral type, being
more significant for early types; this effect might contribute to the 
discrepancy between different observational results, and should be taken 
into account when selecting the local reference TF sample.

%
\section*{Acknowledgements}
This study has been financed by the Academy of Finland (grant nr.~208792),
by a EU Marie Curie Intra-European Fellowship under contract 
MEIF-CT-2005-010884,
and by the Danmarks Grundsforskningsfond through the establishment of the
(now expired) Theoretical Astrophysics Center and of the DARK Cosmology
Centre. All computations reported in this paper were performed on the IBM 
SP4 and SGI Itanium II facilities provided by Danish Center for Scientific 
Computing (DCSC). 

LP acknowledges kind hospitality from the Astronomical Observatory 
and from DARK in Copenhagen on various visits, and useful discussions 
with Frank van den Bosch, Steven Bamford and Stacy McGaugh.

\end{document}